\documentclass[conference]{IEEEtran}

\AtBeginDocument{%
  }
\usepackage{multirow}
\usepackage{subcaption}
\usepackage{amsmath}
\usepackage{enumitem}
\usepackage{graphicx}
\usepackage{booktabs}
\usepackage{xcolor}
\usepackage{amsfonts}
\usepackage{xspace}

\DeclareFontFamily{U}{mathb}{}
\DeclareFontShape{U}{mathb}{m}{n}{
  <-5.5> mathb5
  <5.5-6.5> mathb6
  <6.5-7.5> mathb7
  <7.5-8.5> mathb8
  <8.5-9.5> mathb9
  <9.5-11.5> mathb10
  <11.5-> mathb12
}{}
\DeclareSymbolFont{mathb}{U}{mathb}{m}{n}
\DeclareMathSymbol{\ulsh}{3}{mathb}{"E8}
\DeclareMathSymbol{\ursh}{3}{mathb}{"E9}
\DeclareMathSymbol{\dlsh}{3}{mathb}{"EA}
\DeclareMathSymbol{\drsh}{3}{mathb}{"EB}

\newcommand{\ie}{\emph{i.e.}\xspace}

\newcommand{\Comment}[1]{}

\newcommand{\APN}{GLSTaGAT}

\begin{document}

\title{Global-local Spatial-temporal Aware Graph Attention Network for Network Traffic Forecasting}

\author{\IEEEauthorblockN{1\textsuperscript{st} Jinming Xing}
\IEEEauthorblockA{\textit{North Carolina State University} \\
jxing6@ncsu.edu}
\and
\IEEEauthorblockN{2\textsuperscript{nd} Guoheng Sun}
\IEEEauthorblockA{\textit{University of Mayland} \\
ghsun@umd.edu}
\and
\IEEEauthorblockN{3\textsuperscript{rd} Hui Sun}
\IEEEauthorblockA{\textit{North Carolina State University} \\
hsun26@ncsu.edu}
\and
\IEEEauthorblockN{4\textsuperscript{th} Lichao Pan}
\IEEEauthorblockA{\textit{Shenzhen University} \\
panlinchao2019@email.szu.edu.cn}
\and
\IEEEauthorblockN{5\textsuperscript{th} Shakir Mahmood}
\IEEEauthorblockA{\textit{North Carolina State University} \\
smahmoo2@ncsu.edu}
\and
\IEEEauthorblockN{6\textsuperscript{th} Xuanhao Luo}
\IEEEauthorblockA{\textit{North Carolina State University} \\
xluo26@ncsu.edu}
\and
\IEEEauthorblockN{7\textsuperscript{th} Muhammad Shahzad}
\IEEEauthorblockA{\textit{North Carolina State University} \\
mshahza@ncsu.edu}
}

\maketitle

{
\sloppy
\begin{abstract}
    Spatial-temporal network traffic forecasting is a challenging task due to the complex spatial relationships and dynamic temporal patterns present in each node. Traditional regression methods are not directly applicable to such graph data. Recently, Graph Neural Networks (GNNs) have been widely used to model spatial-temporal dependencies. However, existing methods face several limitations: (1) They rely solely on a predefined spatial adjacency matrix, overlooking hidden low-level temporal information. (2) They model spatial and temporal information separately, which inevitably leads to a loss of joint dependencies, or they capture only global or local dependencies. To address these issues, we propose the \textbf{G}lobal-\textbf{L}ocal \textbf{S}patial-\textbf{T}emporal \textbf{a}ware \textbf{G}raph \textbf{AT}tention Network (GLSTaGAT). Specifically, we adopt a data-driven spatial-temporal fusion graph that incorporates low-level spatial and temporal information, serving as the foundation for further graph convolutions. The GLSTaGAT block and its pooling variant are proposed to simultaneously capture local and global spatial-temporal dependencies. Additionally, we introduce node normalization to mitigate covariance shifts, enabling a smoother training process. An encoder-only transformer is utilized to model high-level joint dependencies, and a multi-head attention prediction layer is designed for final information aggregation and prediction. Experimental results on real-world datasets demonstrate that GLSTaGAT outperforms the baselines by 32.14\% (MAE), 28.30\% (RMSE), and 20.47\% (SMAPE) on average.

\end{abstract}

\vspace{-0.08in}
\section{Introduction}
\label{sec:Introduction}
\vspace{-0.02in}

\noindent\textbf{Motivation.}
In today's complex and rapidly growing networked systems, accurate prediction of various aspects of network traffic is essential for performing a variety of critical tasks, such as optimizing resource allocation, mitigating congestion, meeting service level agreements, and many more.
The traffic dynamics in these networked systems are inherently spatio-temporal, with traffic patterns influenced by the interactions among nodes as well as by the temporal fluctuations in the demands of the nodes.
These interactions and fluctuations further exist at both local and global levels in both spatial and temporal domains. 
The global spatial dependencies encompass interactions across multiple or all nodes while local spatial dependencies encompass interactions among neighboring nodes. 
Similarly, global temporal dependencies exist due to patterns that span over long durations in time-series of measurements while local temporal dependencies come from short lived patterns, such as abrupt spikes, drops, or transient trends.
Thus, there is a growing need for an approach that simultaneously considers the spatial and temporal aspects of network at both local and global scales to predict any desired network traffic metrics.
These metrics include, but are not limited to, traffic volume, number of flows, throughput, latency, and so on.

\vspace{0.03in}
\noindent\textbf{Limitation of Prior Art.}
%
%
Traditional statistical prediction methods, such as ARIMA \cite{kumar2015short}, and time-series forecasting techniques, such as SVR \cite{castro2009online} are no longer able to predict network traffic metrics because the temporal and spatial dependencies in traffic metrics at various nodes in today's networks are no longer linear.
Unfortunately, these approaches struggle when capturing any non-linear and dynamic dependencies.

Deep learning based models have shown promise in overcoming some of these challenges, but still fail to fully exploit the spatio-temporal relationships \cite{li2021spatial}.
This has led to an increased interest in using graph neural networks (GNNs) \cite{wu2022graph}, which are naturally suited to handle graph-structured data.
Although advances have been made in GNNs to exploit spatio-temporal dependencies, two key limitations remain. 

The first limitation is that most prior GNN-based approaches capture only the spatial proximity of the nodes but do not take into account temporal similarity among them \cite{li2021spatial}.
The few prior approaches that have attempted to incorporate temporal as well as spatial dependencies, handle them separately, resulting in unnecessary transformations between semantic spaces 
\cite{li2017diffusion,zhao2019t,bai2021a3t,wu2019graph,wang2020traffic}.
The net effect is similar to that of running two predictors in parallel, one that captures spatial information and the other that captures temporal information, and combining their outputs at the end. 
This inevitably results in not capturing the joint spatio-temporal dependencies, which are crucial for accurate prediction of traffic metrics.

The second limitation is that none of the prior approaches capture global and local dependencies simultaneously for both spatial and temporal domains.
Some approaches, such as \cite{wang2024fully,wang2020traffic,wu2019graph}, have attempted to capture both spatial and temporal dependencies, but only from the local or the global perspective; not from both, simultaneously.
We show in this paper that there is a significant amount of information to benefit from when spatial and temporal dependencies are jointly considered at both local and global scales.

\vspace{0.03in}
\noindent\textbf{Problem Statement.}
Our goal is to develop an approach, which, when given a time-series of measurements of a specific network metric, collected at various nodes in a network, can accurately predict the future values of that metric at those same nodes.
The approach should achieve this by learning joint spatio-temporal dependencies simultaneously at both global as well as local scales.

\vspace{0.03in}
\noindent\textbf{Proposed Approach.}
\label{sec:ProposedApproach}
In this paper, we propose \APN, which achieves the goal stated above.
Fig. \ref{fig:Framework} shows the block diagram of \APN.
\APN~performs the prediction in four steps.
In the first step, it creates a spatio-temporal representation of the network topology that represents not only the spatial proximity of the nodes with each other but also the temporal proximity in their traffic patterns.
In the second step, \APN~takes this representation, along with the actual traffic measurements, and provides them to a spatial modeling block.
The spatial modeling block uses graph attention network (GAT) as its foundational structure and creates a representation that models all local spatial dependencies in the traffic measurements.
\APN~further performs pooling on this representation to capture the global spatial dependencies in the traffic measurements.
In the third step, \APN~provides the output of the spatial modeling block and of the pooling process to two independent encoder-only transformer blocks to capture both global and local temporal dependencies.
%
%
\APN~then combines the outputs of the two transformer blocks, which results in a representation of the input data that simultaneously models both spatial and temporal dependencies at both global and local levels, and can now predict future traffic measurements.
In the last step, \APN~passes this representation through the prediction block, which employs a multi-head attention mechanism to further model any sequential dependencies, and generates predictions for future traffic measurements.
%

\vspace{0.03in}
\noindent\textbf{Technical Challenges.}
In designing \APN, we faced several technical challenges, of which we mention a few here and remaining in the main body of the paper.
The first challenge was to quantify the temporal proximity among nodes and represent it in the adjacency matrix (adjacency matrices have conventionally only been used to represent the spatial proximity among nodes).
To address this challenge, we employ a data-driven approach based on dynamic time warping \cite{li2021spatial} to first quantify the temporal proximity among all pairs of nodes and then add that information into the spatial adjacency matrix to obtain a spatio-temporal adjacency matrix that represents both spatial as well as temporal proximity among nodes.
The second challenge was to mitigate covariance shifts, which make it difficult for the model to converge.
To address this challenge, we introduce a novel technique, Node Normalization, which mitigates covariance shifts by performing normalization within each node across both the time and feature dimensions.
The third challenge was to capture \emph{joint} spatio-temporal dependencies (unlike prior works, which handle temporal and spatial aspects of network measurements separately).
To address this challenge, we first obtain representations that capture local and global spatial dependencies and then apply encoder-only transformers on these representations instead of on the raw traffic measurements, enabling \APN~to capture the spatial and temporal dependencies simultaneously.
%
%
Other significant challenges included capturing global and local spatial dependencies simultaneously, which we addressed by incorporating a pooling process, increasing the prediction horizon, which we addressed by employing an attention mechanism in the prediction layer, and several more.

\vspace{0.03in}
\noindent\textbf{Key Contributions.}
To summarize, in this paper, we make five key contributions.
We present:

\begin{enumerate}[leftmargin=*]
\item a novel approach that simultaneously models spatio-temporal dependencies in network measurements across multiple nodes at both global and local scales.
\item a unique way to represent both spatial and temporal proximity in a single adjacency matrix.
\item a simple yet effective learnable pooling mechanism to capture global spatial trends.
\item a normalization technique for spatio-temporal graphs to mitigate covariance shifts.
\item extensive evaluation and comparison of \APN~with prior approaches using data collected from two large real-world networks. Results show that \APN~improves the mean absolute error, the root mean square error, and the symmetric mean absolute percentage error by 32.14\%, 28.30\%, and 20.47\%, respectively.
%
\end{enumerate}
We will release the source code and data used in the evaluation and comparison of \APN~on the acceptance of this paper.

\begin{figure*}[htbp]
  \centering
  \includegraphics[width=0.95\linewidth]{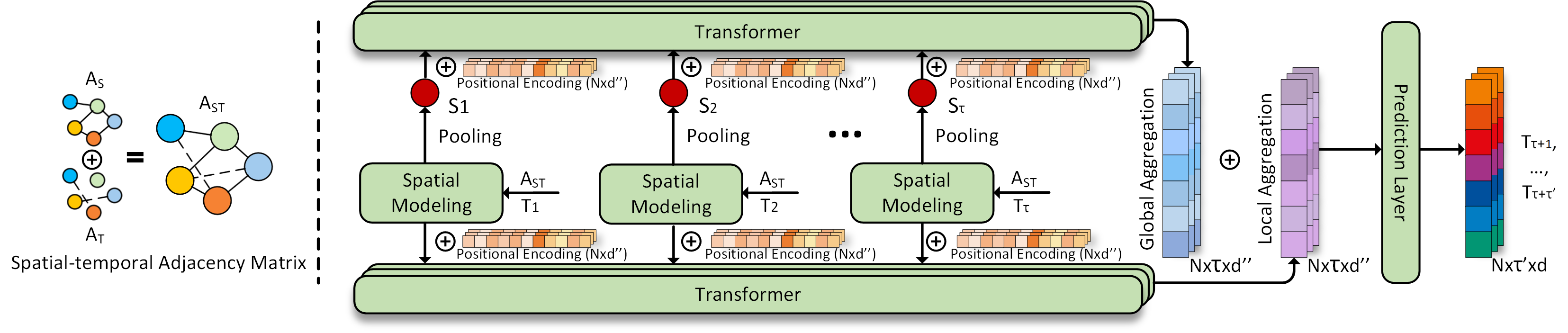}
\vspace{-0.1in}
  \caption{Block diagram of the \APN~framework.}
    \vspace{-0.2in}
  \label{fig:Framework}
\end{figure*}
\vspace{-0.1in}

\section{Related Work}
We broadly categorize prior approaches into three classes: statistical, deep learning (DL) based, and GNN-based.

\noindent\textbf{Statistical Approaches}.
Before DL, statistical models were widely used.
Linear regression \cite{sun2002short} and its numerous variants \cite{navarro2022traffic}, vector auto regressions \cite{lippi2013short}, and support vector regression \cite{castro2009online,li2021application} are among the most prominent.
Unfortunately, these methods often have limited applicability on today's real-world data as they require the traffic time-series to be stationary.
Although ARIMA \cite{kumar2015short} was introduced to handle non-stationary scenarios, it still struggles with turning points and fails to learn periodic patterns.
Recently, Amiri et. al., \cite{amiri2023ensemble} proposed STK-EBM, an ensemble-based method that  utilizes random forest, $k$-nearest neighbor, and XGBoost, to achieve better results compared to prior approaches.
Unfortunately, these statistical methods are not capable of incorporating spatial information across nodes, and work independently on single streams of data.

\noindent\textbf{DL-based Approaches.}
Feed-forward neural networks \cite{vaswani2017attention} were directly applied to time-series, but they are prone to over-fitting and difficult to train.
Recurrent neural networks \cite{grossberg2013recurrent} and their successors, such as long-short-term memory \cite{graves2012long} and gated recurrent units (GRU) \cite{chung2014empirical}, were developed to handle sequential data.
These models significantly outperform statistical methods due to their ability to learn complex patterns from large amounts of data.
Transformers with self-attention \cite{vaswani2017attention} and their variants \cite{lin2022survey} have proven even more effective in modeling sequential data, offering improvements in accuracy \cite{he2024matters}.
Recently, Ma et. al., \cite{ma2023cellular} proposed a correlation-based convolutional LSTM and self-attention-based network to predict complex cellular network traffic accurately.
Unfortunately, none of these prior approaches handle spatial information due to inherent limitation in the nature of their design.

\noindent\textbf{GNN-based Approaches.}
GNNs have been widely used to model data with graph structures and non-Euclidean relationships \cite{zhou2020graph,yang2023explanation}.
GNNs can generally be divided into recurrent GNNs \cite{chen2019gated}, convolutional GNNs \cite{kipf2016semi}, graph autoencoders (GAE) \cite{pan2018adversarially}, and spatial temporal GNNs (STGNN) \cite{liu2024spatial,song2020spatial,li2021spatial}.
These models have been applied in a variety of tasks, including node classification, graph classification \cite{huang2024cost}, traffic forecasting \cite{long2024unveiling}, link prediction \cite{zhang2018link}, knowledge graphs \cite{zhang2020relational}, community detection \cite{chen2017supervised}, and recommendation systems \cite{wu2022graphrecommender}.
%
%
TGCN \cite{zhao2019t} is one of the pioneering works.
It combines graph convolutional networks (GCNs) with GRUs to capture spatio-temporal dependencies.
Attention mechanisms were introduced to enhance TGCN in \cite{bai2021a3t}.
DCRNN \cite{li2017diffusion} models spatial dependencies through random walks and adopts an encoder-decoder architecture to capture temporal dependencies.
GraphWaveNet \cite{wu2019graph} captures both local and global spatial information hierarchically through stacked one-dimensional dilated convolutions.

Recently, STFGNN \cite{li2021spatial} proposed a spatio-temporal fusion graph for better joint dependency understanding, while STGNN \cite{wang2020traffic} leveraged GCNs and transformers for both local and global aggregation.
In \cite{liu2024spatial}, STECAGCN incorporated reinforcement learning with spatio-temporal GNNs for both traffic prediction and load balancing.
Jiang et al. \cite{jiang2024mobile} proposed a multimodal GNN model that fuses information from SMS, call, and Internet services.
In \cite{gupta2023frigate}, gated Lipschitz embeddings and LSTMs are combined into GNNs for information fusion, while Wang et al. \cite{wang2024fully} explored fully connected graph construction and convolution.
In \cite{geng2024stgaformer}, gated self-attention is utilized for both spatial and temporal dependency modeling.
Despite the remarkable achievements of these methods, they treat spatial and temporal information separately and rely entirely on the given adjacency matrix, which limits their ability to handle dynamic and complex spatial dependencies simultaneously with temporal dependencies.
This limitation, in turn, keeps them from achieving the true potential of traffic prediction accuracy with GNNs, as we will show during our evaluation section.
%
We strive to overcome these limitations in our design of \APN.

\vspace{-0.05in}
\section{Formal Problem Definition}
\label{sec:ProblemStatement}
Consider a network consisting of $N$ nodes, $n_1, n_2, \ldots,n_N$, connected to each other through some topology.
We use the adjacency matrix $A_S$ of dimensions $N \times N$ to specify this topology.
Any element $a_{ij}$ of this adjacency matrix is equal to 1 if nodes $n_i$ and $n_j$ are direct neighbors, and 0 otherwise.
We assume that we have historical traffic measurements for the last $\mathcal{\tau}$ measurement intervals from all $N$ nodes.
Let $T_i\in \mathbb{R}^{N \times d}$ (\ie, $T_i$ is a matrix of real numbers with dimensions $N\times d$) represent the traffic measurements obtained from all $N$ nodes at measurement interval $i$.
%
%
Although $d$ typically equals 2 to incorporate measurements of both inbound and outbound traffic, the design of \APN~is generic and can handle any value of $d$.
Let $\mathcal{T}\in\mathbb{R}^{\tau\times N \times d}$ be the matrix obtained from stacking all $T_i$, \ie, $\mathcal{T}=\{T_1; T_2; \dots; T_\tau\}$.
In other words, $\mathcal{T}$ represents historical traffic measurements from all $N$ nodes at all $\tau$ measurement intervals.
Our goal is to construct a model $\mathcal{F}$ that takes an adjacency matrix $A$ and the historical traffic sequence $\mathcal{T}$ as input and outputs predicted traffic measurements $\mathcal{\hat{T}} = \{T_{\tau+1}, T_{\tau+2}, \dots, T_{\tau+\tau'}\}$ for the next $\tau'$ intervals.
This forecasting problem is formally expressed as:
\vspace{-0.05in}
\begin{equation}
\vspace{-0.05in}
  \mathcal{\hat{T}} = \mathcal{F}(A, \mathcal{T})
\end{equation}
Before proceeding, we would like to clarify that the letters $i$, $j$, $k$, and $l$ do not have a fixed meaning in this paper. 
We simply use them either as iteration variables or to represent a specific instance of another defined metric used in this paper.

\section{Proposed Framework}
We provided the overview of \APN~in Sec. \ref{sec:Introduction}, where we described the four steps that constitute \APN.
In the next four subsections, we provide detailed descriptions of these four steps along with the various blocks shown in Fig. \ref{fig:Framework}.

\subsection{Spatio-Temporal Adjacency Matrix}
\label{subsec:Spatio-temporalAdjacencyMatrix}
The adjacency matrix $A_S$ only quantifies the spatial proximity of the nodes but not the temporal similarity in their traffic patterns.
Often, nodes with similar temporal traffic patterns are not one-hop neighbors.
Our goal is to first compute a temporal adjacency matrix, $A_T$, which quantifies the temporal similarity between the traffic patterns of any pair of nodes and then combine $A_S$ and $A_T$ to obtain matrix $A_{ST}$ that simultaneously quantifies spatial and temporal proximity between any pair of nodes.

To measure the similarity between traffic patterns of any pair of nodes, \APN~uses dynamic time warping (DTW) \cite{li2021time}, a well-known method to quantify the similarity between two sequences.
As the computational complexity of the conventional DTW is $O(N^2)$, in \APN~, we instead use fast DTW \cite{li2021spatial}.
Fast DTW restricts the warping path within the search range by setting a constant search length $L$, reducing the time complexity to $O(LN)$, making it much more feasible for use with large amounts of data.
Let $U_j \in \mathbb{R}^{\tau \times d}$ represent the traffic measurements at node $n_j$ in the $\mathcal{\tau}$ most recent measurement intervals.
As each measurement has $d$ values, \ie, $d$ dimensions, and as DTW works only with one dimensional sequences, to calculate the dissimilarity score $D_{jk}$ between any pair of nodes $n_j$ and $n_k$, \APN~takes one dimension at a time, applies DTW to the measurement sequences of the two nodes for that dimension, and then averages the $d$ DTW values to obtain the final value of $D_{jk}$.
As DTW gives a dissimilarity matrix, we convert that into a similarity matrix by simply taking the inverse of each $D_{jk}$.
Thus, the element in the $j^{\text{th}}$ row and the $k^{\text{th}}$ column of $A_T$ is equal to $1/D_{jk}$.

Next, we add $A_S$ to $A_T$ to obtain the spatio-temporal adjacency matrix $A_{ST}$.
It is appropriate to add $A_S$ to $A_T$ because both matrices contain information about the same quantity, \ie, proximity: $A_S$ quantifies spatial proximity among nodes, while $A_T$ quantifies temporal proximity among nodes.
We observed that $A_{ST}$ obtained this way resulted in a dense matrix, containing noise, which impairs the performance of graph convolutions.
To address this, \APN~applies a filter where it sets all entries in $A_{ST}$ that fall in the top-$p$ percentile as one and makes all remaining entries zero.
The resulting adjacency matrix, $A_{ST}$, is a binary matrix that not only quantifies spatial proximity among nodes but also temporal proximity, enabling \APN~to capture spatio-temporal dependencies among nodes.

The value of $p$ in the top-$p$ percentile plays a critical role in establishing a trade-off between information loss and noise.
A smaller value of $p$ reduces noise, but comes at the risk of losing information about joint spatio-temporal dependencies.
In contrast, a larger value of $p$ provides richer spatio-temporal context, which can improve prediction accuracy, but 
a very large value of $p$ increases the risk of incorporating noise, which hinders convergence and can actually degrade the prediction accuracy.
We will investigate the impact of the value of $p$ in the evaluation section and provide practical guidance to appropriately set it.

\subsection{Spatial Dependency Modeling}
\label{subsec:SpatialDependencyModeling}
When modeling spatial relationship between measurements at various nodes, \APN~must learn both local dependencies, which various nodes within small clusters have, as well as global dependencies, which the nodes that are not in close proximity still exhibit.
Next, we first describe how \APN~models local dependencies and after that explain how it models the global dependencies.

\begin{figure}[htbp]
\centering
\includegraphics[width=\linewidth]{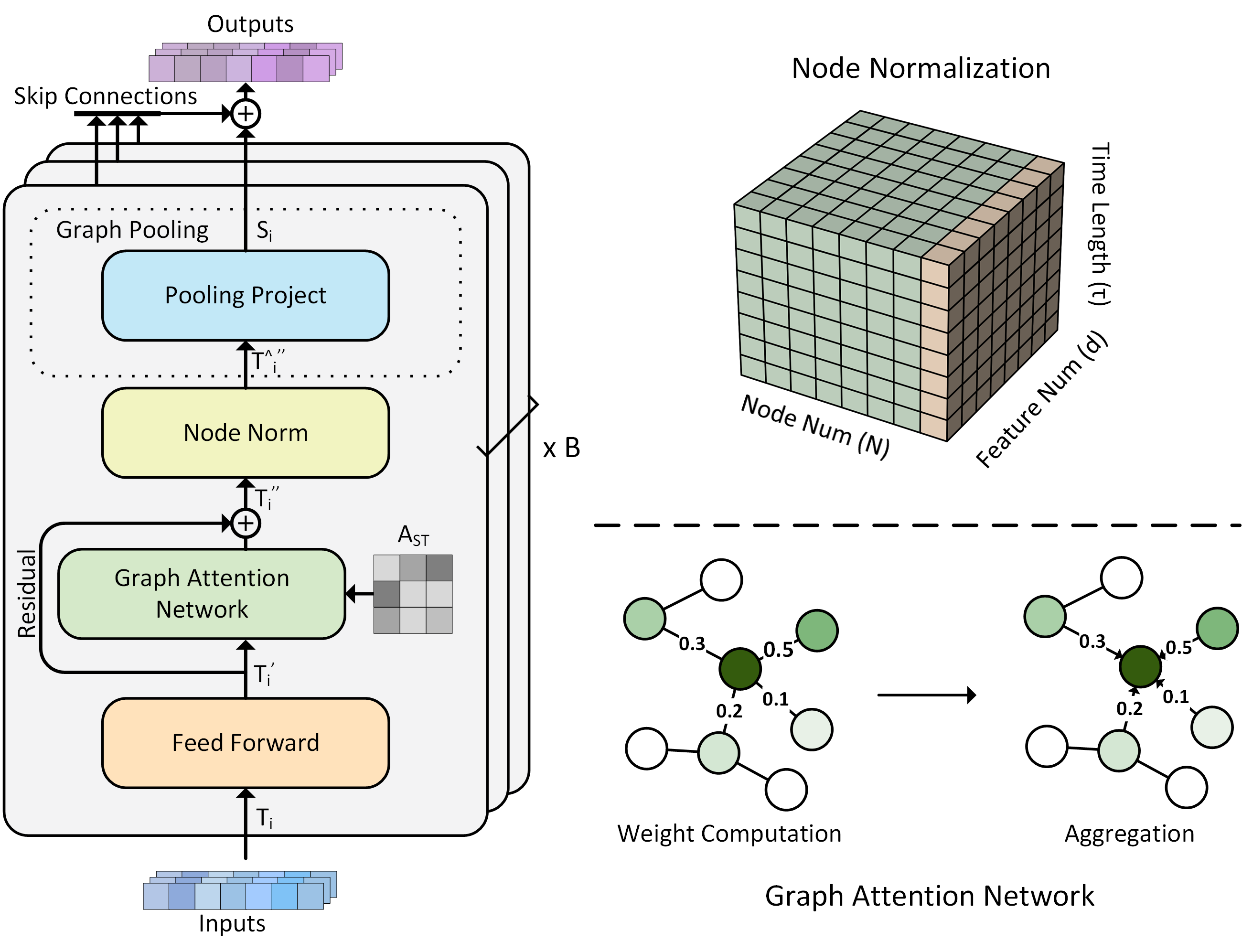}
\vspace{-0.2in}
\caption{Modeling local and global spatial dependencies.}
\vspace{-0.1in}
\label{fig:NetSight_Block}
\end{figure}

\vspace{0.03in}
\subsubsection{Local Spatial Dependency Modeling}
\label{subsubsec:LocalSpatialDependencyModeling}
Fig. \ref{fig:NetSight_Block} shows the block diagram of the approach that \APN~uses to model local spatial dependencies.
It consists of three components: a feed-forward layer, a graph attention layer, and a node normalization layer.

\vspace{0.04in}
\noindent\textbf{Feed-forward Layer:}
The feed-forward layer is used for feature aggregation and transformation.
Recall that $T_i\in \mathbb{R}^{N \times d}$ represents the traffic measurements obtained from the $N$ nodes during the $i^\text{th}$ measurement interval,
Given the traffic measurements $T_i$, the feed-forward layer produces an output $T_i' \in \mathbb{R}^{N \times d'}$ according to the following equation:
\begin{equation}
T_i' = \sigma(T_i\times W_\text{FF} \oplus b)
\label{eq:feedForward}
\end{equation}
where $W_\text{FF} \in \mathbb{R}^{d \times d'}$ and $b \in \mathbb{R}^{d'}$ are learnable parameters, $\sigma$ represents the ReLU activation function, and $\oplus$ represents the broadcast addition.
We represent the $d\times 1$ vector measured at any node $n_j$ during the $i^\text{th}$ measurement interval with $T_i[j]$.
Similarly, we use $T_i'[j]$ to represent the $d'\times1$ vector corresponding to this node $n_j$ after $T_i$ has passed through the feed-forward layer. 
These two notations will be extensively used in the rest of this section.

\vspace{0.04in}
\noindent\textbf{Graph Attention Layer:}
This layer captures the local spatial dependencies between the nodes.
To achieve this, we use a graph attention network (GAT) because its attention-based aggregation enables a versatile message passing and local information capture.
As a first step, for any given node $n_j$, \APN~calculates the attention score $e_{jk}$ between that node $n_j$ and its immediate neighbor $n_k$ through a linear transformation $a: \mathbb{R}^{d'} \times \mathbb{R}^{d'} \to \mathbb{R}$.
This linear layer transforms the data to a dimension that is more suitable to calculate the relationships among neighbors.
Formally
\begin{equation}
e_{jk} = a\Big(W_a\times T_i'[j]^\mathbb{T}, W_a\times T_i'[k]^\mathbb{T}\Big)
\end{equation}
where $W_a\in \mathbb{R}^{d\times d'}$ is a learnable parameter and $\mathbb{T}$ represents the matrix transpose operation.
The attention score $e_{ij}$ is a scaler that quantifies the importance of node $j$ to node $i$.
Next, \APN~applies softmax normalization on all attention scores, $e_{jk}$, as specified by the following equation.
\begin{equation}
  \alpha_{jk} = \text{softmax}(e_{jk}) = \frac{\exp(e_{jk})}{\sum_{\forall l \in \mathcal{N}(n_j)} \exp(e_{jl})}
\label{eq:softmax}
\end{equation}
where $\mathcal{N}(n_j)$ represents the set of all the immediate neighbors of the node $n_j$, as specified in the spatio-temporal adjacency matrix $A_{ST}$.

Next, \APN~passes $T_i'$ through another linear layer
to capture local dependencies between the traffic measurements from neighboring nodes.
We represent the embedding for node $n_j$ after passing $T_i'[j]$ through this linear layer with $T_i''[j] \in \mathbb{R}^{1 \times d''}$, which is computed as:

\noindent
\begin{equation}
  T_i''[j] = \sigma\left(\sum_{l \in \mathcal{N}(n_j)} \alpha_{jl} \Big(T_i'[l]\times W_G\Big) \right)
\end{equation}
where $W_G\in \mathbb{R}^{d'\times d''}$ is also a learnable parameter.
As noted in \cite{vaswani2017attention}, if instead of using the single instance of the linear layer described above, we employ a multi-head attention mechanism, \ie, if we use multiple independent parallel linear layers and aggregate their output using an appropriate approach, the resulting embedding would capture spatial information from multiple perspectives.
In \APN~we employ this multi-head attention mechanism and aggregate their output as described in the equation below.
\begin{equation}
  T_i''[j] = \sigma\left(\frac{1}{M} \sum_{m=1}^{M}\sum_{l \in \mathcal{N}(n_j)} \alpha_{jl}^m \Big(T_i'[l]\times W_G^m\Big) \right)\\
\end{equation}
where $\alpha_{jl}^m$ represents the attention weight of node $l$ to node $j$ in the $m^\text{th}$ head, $W_G^m\in \mathbb{R}^{d'\times d''}$ represents the transformation matrix for the $m^\text{th}$ head, and $M$ represents the number of heads.
This embedding $T_i''[j]$ of node $n_j$ captures any underlying relationships that the traffic measurements at node $n_j$ have with the traffic measurements at the neighbors of $n_j$, thus modeling the local dependencies at each node in the network.

\vspace{0.04in}
\noindent\textbf{Normalization Layer:}
It is well-known that embeddings, such as $T_i''[j]$, experience covariance shifts.
To mitigate these covariance shifts and to expedite the training process, two common approaches that have been proposed in literature for convolutional neural networks (CNNs) and recurrent neural networks (RNNs) are batch  normalization \cite{santurkar2018does} and layer normalization \cite{xu2019understanding}.
While these approaches are suitable for CNNs and RNNs, unfortunately, when used in graph neural networks, such as the one we are working with, batch  normalization only normalizes along the node dimension and layer normalization only normalizes on the feature dimension, and thus do not facilitate the joint learning of spatio-temporal dependencies.
Consequently, we develop a new normalization approach for use in \APN, namely node normalization, which mitigates covariance shifts by performing normalization within each node across both the time and feature dimensions, and thus
achieving our goal of joint spatio-temporal learning.

Recall that the subscript $i$ in $T_i''[j]$ represents the $i^\text{th}$ measurement interval.
Furthermore, recall that $T_i''[j]$ is a vector of dimensions $1\times d''$.
Let us represent the $k^\text{th}$ value of this vector with $T_i''[j][k]$, where $1\leq k\leq d''$.
%
%
To perform node normalization for any node $n_j$, \APN~first calculates the mean and standard deviations using the following equations:

\vspace{-0.05in}
\begin{equation}
\mu_j = \frac{1}{T\times d''}\sum_{i=1}^T\sum_{k=1}^{d''} T_i''[j][k]
\end{equation}
\vspace{-0.05in}
\begin{equation}
\sigma_j = \left(\frac{1}{T\times d''}\sum_{i=1}^T\sum_{k=1}^{d''} \left(T_i''[j][k]-\mu_j\right)^2\right)^{1/2}
\end{equation}
Next, \APN~scales all $T_i''[j][k]$ values for each node $n_j$ using the corresponding $\mu_j$ and $\sigma_j$ values as well as using two learnable parameters, $\hat{\mu}_j$ and $\hat{\sigma}_j$,
as shown below.

\noindent
\begin{equation}
\forall_{i=1}^{T},\forall_{j=1}^{N},\forall_{k=1}^{d''},\quad \hat{T}_i''[j][k] = \hat{\sigma_j} \times \frac{T_i''[j][k] - \mu_j}{\sigma_j} + \hat{\mu_j}
\label{eq:nodeNormLayer}
\end{equation}

Finally, note from Fig. \ref{fig:NetSight_Block} that there is a residual connection from the feed forward layer that is added to the output of the graph attention layer before the output is fed to the node normalization layer.
We did this to stabilize the training process and to prevent forgetting the prior learned knowledge.

\vspace{0.03in}
\subsubsection{Global Spatial Dependency Modeling}
\label{subsubsec:GlobalSpatialDependencyModeling}
To model the global spatial dependencies, which exist throughout all nodes in the topology or at least exist among cluster of nodes  larger than just the immediate neighbors, we employ a pooling mechanism.
While some global pooling methods have been proposed in prior literature, such as mean and max pooling \cite{zhao2024improved}, and hierarchical pooling methods such as gPool \cite{lee2019self} and DiffPool \cite{ying2018hierarchical}, they are not well-suited for our problem because these methods either lack sufficient learnability or are overly complex for our setting.

We propose a simple, yet effective, pooling mechanism, where \APN~obtains a weighted average of all nodes in the topology by passing their latest representation through a linear layer, and thus capturing the global spatial dependencies.
We call the output of this linear layer a super node.
More specifically, \APN~passes $\hat{T}_i''\in\mathbb{R}^{N \times d''}$, obtained from Eq. \eqref{eq:nodeNormLayer}, through a pooling layer, as shown in Fig. \ref{fig:NetSight_Block}, and obtains a pooling projection, $S_i$, according to the following equation.
%
\begin{equation}
  S_i = \left(\hat{T}_i''^{\mathbb{T}}\times W_P\right)^\mathbb{T}
\end{equation}
where $W_P\in \mathbb{R}^{N \times 1}$ is a learnable attention matrix, and $S_i\in \mathbb{R}^{1 \times d''}$ is the super node that captures the global spatial dependencies in the measurements obtained during the $i^\text{th}$ measurement interval.

\vspace{0.03in}
\subsubsection{Stacking}
\label{subsubsec:StackingandAggregation}
The three layers described in Sec. \ref{subsubsec:LocalSpatialDependencyModeling} (and the pooling layer described in Sec. \ref{subsubsec:GlobalSpatialDependencyModeling} for modeling global spatial dependencies) constitute one \emph{block}.
%
%
To enhance model's expressive power, we stack $B$ such blocks, as shown in Fig. \ref{fig:NetSight_Block}, and connect consecutive blocks with skip connections.
These skip connections, similar to the residual connections, keep the model from forgetting prior learned knowledge.
%
%

\subsection{Temporal Dependency Modeling}
\label{subsec:Temporal Dependency Modeling}
%
Similar to spatial dependencies, when modeling relationship between measurements across time, \APN~must learn both local and global temporal dependencies simultaneously.
In temporal context, local dependencies constitute patterns that emerge for short durations, such as abrupt spikes, drops, and other such transient trends, whereas global dependencies constitute patterns that span for longer durations.
Recall from Sec. \ref{subsec:Spatio-temporalAdjacencyMatrix} that the spatio-temporal adjacency matrix $A_{ST}$ already incorporates some temporal information.
\APN~captures further temporal dependencies by employing a multi-head attention based transformer encoder.
Fig. \ref{fig:Framework} shows the placement of our transformer in the entire architecture of \APN, and Fig. \ref{fig:TE-MHA} shows the internal architecture of the transformer.
We emphasize here that our transformer encoder captures both local and global temporal dependencies simultaneously, and not in two separate steps, as was the case when we were capturing local and global spatial dependencies in Sec. \ref{subsubsec:LocalSpatialDependencyModeling} and \ref{subsubsec:GlobalSpatialDependencyModeling}, respectively.
\begin{figure}[htbp]
  \centering
\vspace{-0.1in}
  \includegraphics[width=\linewidth]{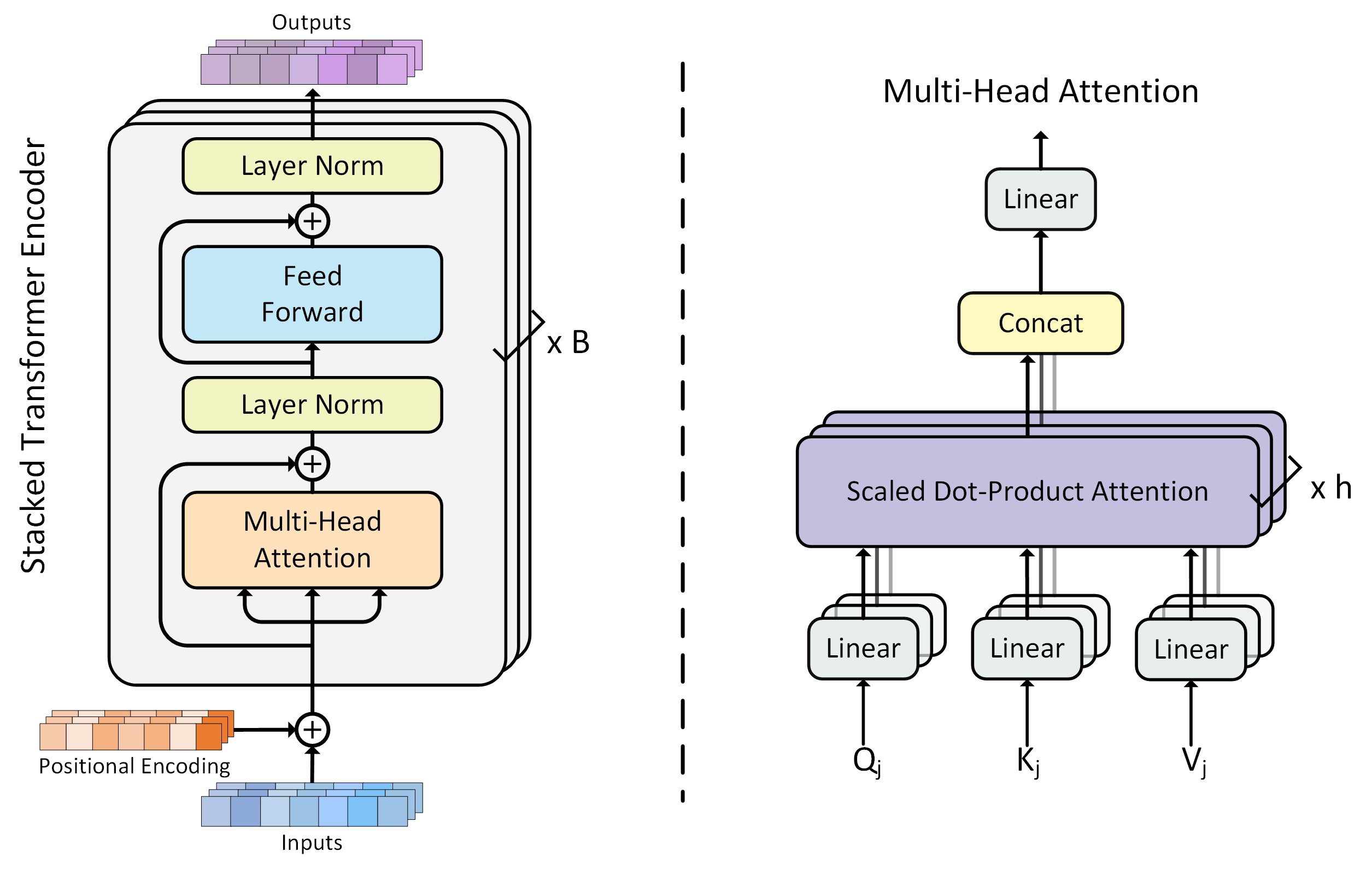}
\vspace{-0.2in}
  \caption{Transformer encoder and multi-head attention.}
\vspace{-0.1in}
  \label{fig:TE-MHA}
\end{figure}

As shown in Fig. \ref{fig:Framework}, \APN~applies the transformer on the output of the normalization layer (which contains information about local spatial dependencies) and separately on the output of the pooling process (which contains information about global spatial dependencies).
It does not apply the transformer directly to the traffic measurements because that would undermine the goal of modeling the spatial and temporal dependencies jointly.
Next, we describe how \APN~applies transfer on the output of the normalization layer.
The method to apply it on the output of the pooling layer is the same.

Our transformer is comprised of three components, describe next: a multi-head attention layer, two normalization layers, and a feed-forward layer.
It is further prepended with positional encoding to introduce relative sequential information.
%

\noindent\textbf{Positional Encoding:}
Recall from Sec. \ref{subsubsec:LocalSpatialDependencyModeling} that for any measurement interval $i$, the output of the normalization layer is represented by $\hat{T}_i''\in\mathbb{R}^{N\times d''}$.
Let us construct a new representation $U_j\in\mathbb{R}^{\tau\times d''}$ using the $\hat{T}_i''$ values that will make it easy for us to explain the approach that \APN~uses to model the temporal dependencies.
More specifically, the $i^{\text{th}}$ row of the matrix $U_j$ is equal to the $j^{\text{th}}$ row of the matrix $\hat{T}_i''$, where $1\leq i\leq \tau$ and $1\leq j\leq N$.

Given the node representation $U_j$, \APN~applies positional encoding to obtain a new representation, $U'_j$, as below:
\begin{equation}
  \begin{split}
    \forall_{j=1}^{\lceil N/2\rceil},\quad U_{2j-1}'[i] = U_{2j-1}[i] + \cos\left(\frac{t}{10000^{2j/d_{\text{m}}}}\right)\\
    \forall_{j=1}^{\lfloor N/2\rfloor},\quad U_{2j}'[i] = U_{2j}[i] + \sin\left(\frac{t}{10000^{2j/d_{\text{m}}}}\right)
  \end{split}
\end{equation}
where $1\leq i \leq \tau$, $d_{\text{m}}$ is the embedding size, and $10000^{2i/d_{\text{m}}}$ is a scaling factor that varies across dimensions, allowing different frequencies of sine and cosine waves to be applied at different positions in the encoding.
This new representation can now be passed through the transformer encoder to model both local and global temporal dependencies.
We will discuss the value of $d_m$
shortly.

\noindent\textbf{Multi-head Attention Layer:}
We use multi-head attention instead of single-head attention because it allows aggregation of information from multiple perspectives.
Fig. \ref{fig:TE-MHA} shows the architecture of our multi-head attention layer.
To explain how it works, let us first describe the single-head attention mechanism, and then describe how we extend it to multi-head attention.
Input to an attention layer includes queries, keys, and values.
%
$U'_j \in \mathbb{R}^{\tau \times d''}$ represents the position encoded output of the normalization layer for node $n_j$.
For any node $n_j$, \APN~calculates the values of query $Q_j\in\mathbb{R}^{\tau\times d_q}$, key $K_j\in\mathbb{R}^{\tau\times d_k}$, and value$ V_j\in\mathbb{R}^{\tau\times d_v}$ by applying linear transformations to the $U'_j$, as shown in the following equation.
As query is for some key, the dimensions of a query and key are the same, \ie, $d_q=d_k$.
\begin{equation}
  Q_j = U'_j\times W_Q,\quad K_j = U'_j\times W_K,\quad V_j = U'_j\times W_V
\label{eq:QKV_calcs}
\end{equation}
where $W_Q \in \mathbb{R}^{d'' \times d_q}$, $W_K \in \mathbb{R}^{d'' \times d_k}$, and $W_V \in \mathbb{R}^{d'' \times d_v}$ are projection matrices that are learned during training.
When doing the positional encoding discussed earlier, \APN~sets $d_m=d_k$ when calculating the $U'_j$ that is used to calculate $Q_j$ and $K_j$ and sets $d_m=d_v$ when calculating the $U'_j$ that is used to calculate $V_j$ in Eq. \eqref{eq:QKV_calcs}.

For each query, \APN~applies attention to all keys by first calculating the dot products between each query-key pair and then dividing the result by $\sqrt{d_k}$ to address the gradient vanishing problem.
Next, it normalizes the resulting values using the softmax function (defined in Eq. \ref{eq:softmax}) to generate weights.
Finally, it obtains the attention matrix using these weights to aggregate the values $V_j$.
Formally, for any node $n_j$, the attention matrix $\mathcal{A}(n_j)\in\mathbb{R}^{\tau\times d_v}$ is calculated as:
\begin{equation}
\mathcal{A}(n_j) = \text{softmax}\left( \frac{Q_j\times K_j^\mathbb{T}}{\sqrt{d_k}} \right)\times V_j
\label{eq:TemporalAttentionApplication}
\end{equation}

To incorporate multiple heads, \APN~concatenates outputs of multiple individual attention heads.
Let $h$ represent the number of attention heads.
The aggregated multi-head attention matrix $\mathcal{A}^M(n_j)\in\mathbb{R}^{\tau\times d''}$ for node $n_j$ is calculated as:
\begin{equation}
  \mathcal{A}^M (n_j) = \left( {\|}^h_{k=1} \mathcal{A}_k(n_j) \right)\times W_\mathcal{A}
  \label{eq:multiheadattention}
\end{equation}
where $W_\mathcal{A} \in \mathbb{R}^{h d_v \times d''}$ represents the learnable linear projection parameters.
Notice that the dimensions of $\mathcal{A}^M (n_j)$ are the same as the dimension of the matrix $U_j'$ that was input to the transformer encoder.

\noindent\textbf{Normalization Layers:}
\APN~applies layer normalization to the outputs of multi-head attention as well as the feed forward layer to rescale them, which makes the resulting values more amenable for training.
For any input $x$, the layer normalization is performed as $\hat{x}=\gamma\times\frac{x-\mu}{\sigma}+\beta$, where $\gamma$ and $\beta$ are learnable parameters, $\mu$ and $\sigma$ are the mean and variance of $x$.
The input to each normalization layer is aggregated with a residual connection from the previous layer to stabilize the training process and to prevent forgetting the previously learned knowledge.

\noindent\textbf{Feed-Forward Layer:}
Feed-forward layer is a linear layer, similar to the one described in Eq. \eqref{eq:feedForward}.

\noindent\textbf{Stacking:}
Similar to the approach described in Sec. \ref{subsubsec:StackingandAggregation}, we stack $B$ transformer encoders with skip connections, as shown in Fig. \ref{fig:TE-MHA}, to enhance model's expressive power.

\noindent\textbf{Global and Local Spatio-Temporal Aggregation:}
The methods described in this Sec. \ref{subsec:Temporal Dependency Modeling}, applied on the output of the normalization layer from Sec. \ref{subsec:SpatialDependencyModeling}, result in a $\tau \times d''$ matrix for each node, which captures the local spatio-temporal dependencies.
\APN~applies these methods on the output of the pooling layer from Sec. \ref{subsec:SpatialDependencyModeling} as well in the exact same way, which also results in a $\tau \times d''$ matrix for the super node and captures the global spatio-temporal dependencies.
Finally, for each node $n_j$, \APN~adds the two matrices, as shown in Fig. \ref{fig:Framework}, to get a matrix $G_j\in\mathbb{R}^{\tau \times d''}$, which simultaneously captures global as well as local temporal and spatial dependencies in the traffic measurements at node $n_j$.

\subsection{The Prediction Layer}
%
Our goal is to predict the traffic measurements for the next $\tau'$ measurement intervals.
A simple approach would be pass the matrix $G$ through a few linear layers and directly obtain the predicted traffic measurements for the next $\tau'$ intervals.
%
%
Unfortunately, this approach can not explicitly utilize the sequence structure, which decreases prediction accuracy and increases training difficulty.
%

To overcome the limitations of this direct transformation approach, we introduce a multi-head attention prediction mechanism in \APN, shown in Fig. \ref{fig:PredictionLayer}.
With this mechanism, \APN~sequentially computes a representation for successive measurement intervals from interval $\tau+1$ all the way to the interval $\tau+\tau'$.
Each time it computes a representation for a measurement interval, it passes that representation through a linear layer to obtain accurate prediction of the traffic measurement for that measurement interval.
%
%
We already have $\tau$ rows in matrix $G_j$, \ie, we already have the representation for the $\tau$ measurement intervals.
Every time \APN~computes the representation for the next measurement interval, it appends that representation at the bottom of $G_j$.
Thus, after $k\leq\tau'$ predictions, the size of $G_j$ grows to $\mathbb{R}^{\tau+k\times d''}$.
\begin{figure}[htbp]
  \centering
    \vspace{-0.1in}
    \includegraphics[width=\linewidth]{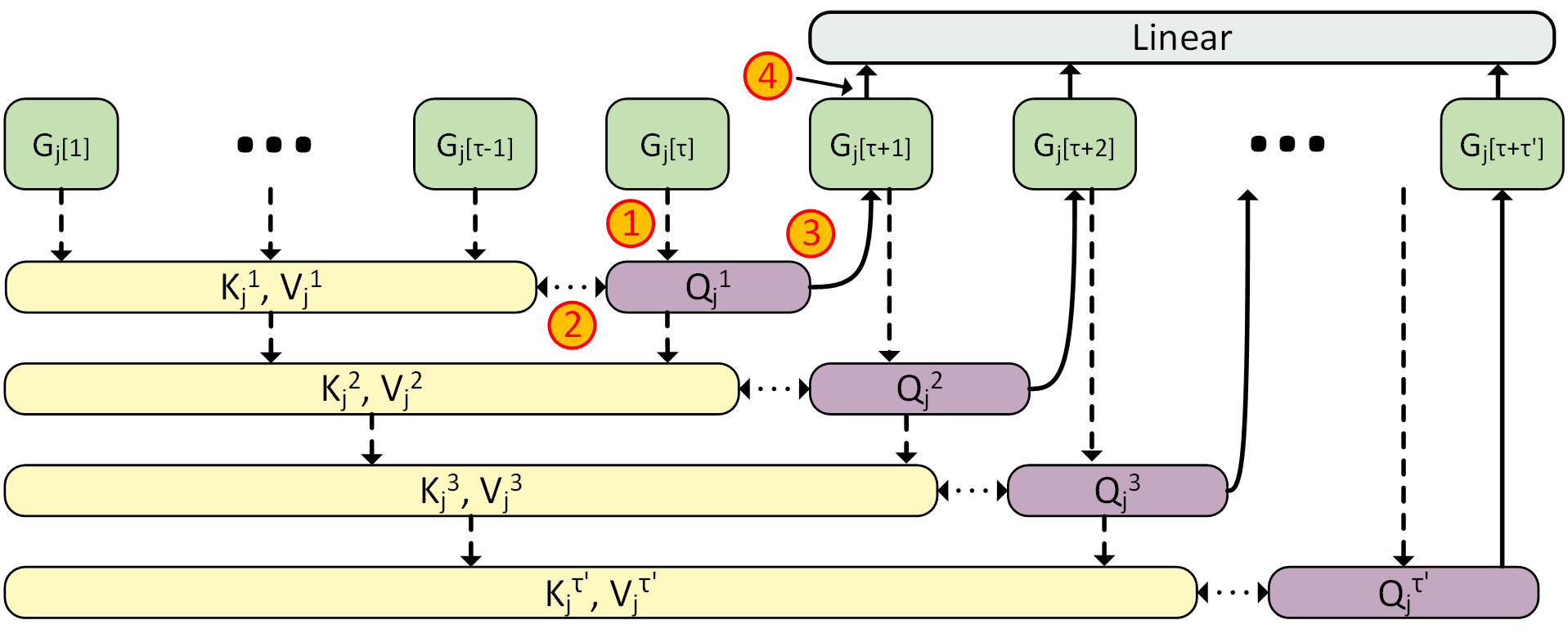}
  \caption{Multi-head attention based prediction layer.}
    \vspace{-0.05in}
    \label{fig:PredictionLayer}
    \end{figure}

Let us use the notation $G_j[1:k]$ to represent the first $k$ rows of the matrix $G_j$, and $G_j[k]$ to represent the $k^{\text{th}}$ row of the matrix $G_j$.
To explain our multi-head attention prediction mechanism, we describe how \APN~computes the $k^{\text{th}}$ representation in future, \ie, for measurement interval $\tau+k$, where $1\leq k\leq\tau'$, and how it uses this computed representation to predict the traffic measurement for this $\tau+k^{\text{th}}$ measurement interval.
For this, \APN~first computes the key $K_j^k=G_j[1:\tau+k-2]\times W_K$ and value $V_j^k=G_j[1:\tau+k-2]\times W_V$, where $K_j^k\in\mathbb{R}^{(\tau+k-2)\times d_k}$ and $V_j^k\in\mathbb{R}^{(\tau+k-2)\times d_v}$.
Next, it computes the query $Q_j^k=G_j[\tau+k-1]\times W_Q$, where $Q_j^k\in\mathbb{R}^{1\times d_k}$.
$W_Q$, $W_K$, and $W_V$ were defined in Eq. \eqref{eq:QKV_calcs}.
\APN~then performs multi-head attention using the computed $Q_j^k, K_j^k, V_j^k$ to obtain $G_j[\tau+k]$, as per the following equation.
\vspace{-0.1in}
\begin{equation}
\vspace{-0.05in}
G_j[\tau+k] = \text{softmax}\left( \frac{Q_j^k\times \left(K_j^k\right)^\mathbb{T}}{\sqrt{d_k}} \right)\times V_j^k
\label{eq:PredictionAttentionApplication}
\end{equation}
where $G_j[\tau+k]\in\mathbb{R}^{1\times d_v}$.
Recall from the paragraph titled ``Normalization Layers'' in Sec. \ref{subsec:Temporal Dependency Modeling} that $d_v=d''$.
Next, it passes $G_j[\tau+k]$ through a linear layer, where $G_j[\tau+k]$ is multiplied with $W_G\in\mathbb{R}^{d''\times d}$.
This results in the final prediction of traffic measurements for the measurement interval $\tau+k$.

\section{Evaluation}
\label{sec:ExperimentalEvaluation}
In this section, we extensively evaluate the performance of \APN.
We first introduce two network traffic datasets collected from two large real-world computer networks and then describe the experimental settings and evaluation metrics.
%
%
After that, we study the performance of \APN~in detail and extensively compare it with 12 state-of-the-art approaches.
Last, we present an ablation study to investigate the impact of the various components of \APN.

\subsection{Datasets}
We chose two well-known real-world network traffic datasets, Abilene \cite{IdzikowskiOrlowskiRaackWoesnerWolisz2010b} and GEANT \cite{KosterKutschkaRaack2010}, described next.

\noindent\textbf{Abilene:}
%
This dataset contains directed network traffic volume across 12 cities in the United States.
The traffic volume is aggregated and averaged every 5 minutes (and reported in Mbits/s), over a period of 167 consecutive days.
Consequently, there were 48,096 measurement intervals, and the dataset contains traffic measurements from all 12 cities (one location per city) in each measurement interval.
%
%

\noindent\textbf{GEANT:}
%
This dataset contains directed network traffic volume across 21 cities in Europe and one city in the United States, aggregated every 15 minutes, over a period of about 4 months.
There were 11,460 measurement intervals, and the dataset contains traffic measurements from all 22 cities (one location per city) in each measurement interval.
%


\subsection{Experimental Settings and Metrics}
\label{subsec:ExperimentalSettingsandMetrics}
For each dataset, we use 70\% of the data for training, 10\% for validation, and the remaining 20\% for testing, as done in a recent work \cite{wang2020traffic}.
We refrained from using cross-validation because it can yield overly optimistic results due to its excessive susceptibility to overfitting.
%
Cross-validation is typically used when the size of the evaluation data set is small.
As our datasets are sufficiently large, to ensure that our results accurately reflect the true performance of our approach, we opted not to use cross-validation.

After splitting each dataset into subsets, we generate sequences using a sliding window of length $\tau + \tau'$, where $\tau$ is the length of the training sequence and $\tau'$ is the length of the forecast sequence.
We set $\tau=12$ while use three different values of $\tau'$, \ie, 6, 12, and 18.
We set the number of layers, hidden size,
and feed-forward size of the transformer to 2, 64, and 128, respectively. 
%
%
We set the maximum number of training epochs to 2000 and employed early stopping and grid search to tune the learning rate, batch size, and number of epochs.
We select the Adam optimizer to ensure smoother training and allow for faster learning.
We adopt the Huber loss function \cite{li2021spatial} because it is less sensitive to outliers compared to the squared error loss.
We set $p=50$ for the top-$p$ percentile.
Finally, we configured all methods that we compare \APN~with using the parameters given in their respective papers.

To quantify the performance of \APN~and prior approaches, we use the three commonly used metrics:
mean absolute error (MAE),
root mean square error (RMSE), and
symmetric mean absolute percentage Error (SMAPE).
Let $T_i[j]$ represent the ground truth measurement at the $j^\text{th}$ node during the $i^\text{th}$ measurement interval.
These three metrics are then defined as:
\vspace{-0.05in}
\begin{equation}
  \text{MAE} = \frac{1}{\tau\times N} \sum_{i=1}^{\tau} \sum_{j=1}^{N} |T_i[j] - \hat{T}_i[j]|
\end{equation}

\vspace{-0.05in}
\begin{equation}
  \text{RMSE} = \left(\frac{1}{\tau\times N} \sum_{i=1}^{\tau}\sum_{j=1}^{N} (T_i[j] - \hat{T}_i[j])^2\right)^{1/2}
\end{equation}

\vspace{-0.05in}
\begin{equation}
  \text{SMAPE} = \frac{1}{\tau\times N} \sum_{i=1}^{\tau} \sum_{j=1}^{N} \frac{|T_i[j] - \hat{T}_i[j]|}{(|T_i[j]| + |\hat{T}_i[j]|)/2}
\vspace{-0.05in}
\end{equation}

\begin{table*}[htbp]
  \centering
  \caption{Traffic prediction on the Abilene dataset.}
\vspace{-0.05in}
  \begin{tabular}{lccccccccc}
    \toprule
    \multirow{2}[3]{*}{Methods} & \multicolumn{3}{c}{30 minutes} & \multicolumn{3}{c}{60 minutes} & \multicolumn{3}{c}{90 minutes}                                                                                                               \\
    \cmidrule(lr){2-4} \cmidrule(lr){5-7} \cmidrule(lr){8-10}
                                & MAE                            & RMSE                           & SMAPE (\%)                     & MAE             & RMSE            & SMAPE (\%)       & MAE             & RMSE            & SMAPE (\%)       \\
    \midrule
    FCLSTM                      & 10.7236                        & 11.8766                        & 35.5436                        & 10.7695         & 11.7635         & 35.9868          & 12.5556         & 13.8763         & 38.7524          \\
    FNN                         & 16.7954                        & 17.4969                        & 42.0241                        & 15.8688         & 16.5912         & 41.7413          & 15.7127         & 16.4054         & 41.6634          \\
    DCRNN                       & 7.7030                         & 8.5114                         & 39.7328                        & 7.5323          & 8.2706          & 39.7013          & 8.3321          & 9.2529          & 40.7549          \\
    TGCN                        & 4.4565                         & \textbf{4.6260}                & 32.0368                        & 4.4804          & \textbf{4.6907} & 32.0911          & 5.4577          & 5.8607          & 33.9991          \\
    GraphWaveNet                & 12.9619                        & 14.8095                        & 45.5251                        & 12.1873         & 13.9285         & 44.1909          & 12.7410         & 14.7815         & 43.4438          \\
    STGNN                       & 8.4696                         & 9.0489                         & 39.0325                        & 8.2686          & 8.8032          & 39.2113          & 8.9264          & 9.5777          & 40.3989          \\
    A3TGCN                      & 10.9353                        & 11.3217                        & 41.8665                        & 12.0779         & 12.9049         & 42.7025          & 14.8635         & 16.1902         & 44.4933          \\
    STFGNN                      & 6.9031                         & 7.1960                         & 49.5838                        & 6.9557          & 7.1673          & 50.0843          & 6.9796          & 7.1685          & 50.2386          \\
    FRIGATE                     & 5.6141                         & 6.5870                         & 32.4451                        & 7.8360          & 8.8796          & 32.2216          & 6.7351          & 6.4902          & 30.4249          \\
    STECAGCN                    & 9.3622                         & 9.9214                         & 36.7264                        & 9.1294          & 9.8064          & 36.1786          & 10.0396         & 10.8432         & 37.1954          \\
    FCSTGNN                     & \textbf{4.4246}                & 5.8367                         & 31.4862                        & 6.6613          & 6.8886          & 31.0963          & 5.3176          & 5.4556          & 32.9388          \\
    STGAFormer                  & 6.8307                         & 5.4176                         & 33.2023                        & 7.3737          & 7.3220          & 34.8981          & 6.1945          & 6.8862          & 31.2072          \\
    \APN                        & 5.2619                         & 6.2408                         & \textbf{30.9784}               & \textbf{4.0905} & 5.0390          & \textbf{28.1796} & \textbf{4.1244} & \textbf{4.9163} & \textbf{28.6257} \\
    \bottomrule
  \end{tabular}%
  \label{tab:Abilene}%
\vspace{-0.1in}
\end{table*}%
\begin{table*}[htbp]
  \centering
  \caption{Traffic prediction on the GEANT dataset.}
\vspace{-0.05in}
  \begin{tabular}{lccccccccc}
    \toprule
    \multirow{2}[3]{*}{Methods} & \multicolumn{3}{c}{90 minutes} & \multicolumn{3}{c}{180 minutes} & \multicolumn{3}{c}{270 minutes}                                                                                                                       \\
    \cmidrule(lr){2-4} \cmidrule(lr){5-7} \cmidrule(lr){8-10}
                                & MAE                            & RMSE                            & SMAPE (\%)                      & MAE               & RMSE              & SMAPE (\%)       & MAE               & RMSE              & SMAPE (\%)       \\
    \midrule
    FCLSTM                      & 476.1998                       & 505.5138                        & 34.3935                         & 527.7311          & 577.4508          & 35.5177          & 584.0931          & 654.8334          & 36.9620          \\
    FNN                         & 560.1513                       & 590.5850                        & 38.4635                         & 581.6550          & 629.8295          & 38.8277          & 614.5461          & 682.0507          & 39.3870          \\
    DCRNN                       & 345.0415                       & 386.6519                        & 29.0962                         & 384.5846          & 441.4062          & 29.5479          & 431.2820          & 507.2540          & 30.6170          \\
    TGCN                        & 267.8098                       & 302.8806                        & 23.0568                         & 350.3870          & 405.9681          & 25.3109          & 433.2575          & 512.2427          & 27.6244          \\
    GraphWaveNet                & 276.0079                       & 315.6213                        & 25.1111                         & 343.2575          & 399.2309          & 26.6782          & 409.6598          & \textbf{483.9348} & 28.4989          \\
    STGNN                       & 459.4285                       & 493.6292                        & 36.5850                         & 491.0048          & 541.5706          & 38.4596          & 538.1622          & 608.9838          & 39.0700          \\
    A3TGCN                      & 458.3262                       & 492.4757                        & 31.6791                         & 485.9204          & 535.2003          & 31.9733          & 519.0844          & 584.8981          & 32.6467          \\
    STFGNN                      & 393.7175                       & 431.0498                        & 32.1377                         & 444.2252          & 499.5881          & 33.6165          & 506.1640          & 580.5396          & 35.5899          \\
    FRIGATE                     & \textbf{236.2362}              & 315.7086                        & 24.9613                         & 361.4566          & 452.4576          & 28.1095          & 449.3897          & 546.1063          & 30.0882          \\
    STECAGCN                    & 399.3089                       & 435.0348                        & 31.8118                         & 446.4186          & 499.8638          & 33.1404          & 502.4053          & 576.1506          & 34.4524          \\
    FCSTGNN                     & 368.4067                       & 415.6187                        & 29.0230                         & 370.2239          & 477.9274          & 29.9675          & 459.5399          & 571.7845          & 33.6143          \\
    STGAFormer                  & 253.7953                       & 302.1323                        & 25.2337                         & 349.2046          & 393.1245          & 26.6804          & \textbf{403.8559} & 498.1085          & 29.0720          \\
    \APN                        & 238.2055                       & \textbf{273.2580}               & \textbf{22.7766}                & \textbf{330.6235} & \textbf{386.9244} & \textbf{25.2505} & 421.8723          & 501.9248          & \textbf{27.5982} \\
    \bottomrule
  \end{tabular}%
  \label{tab:GEANT}%
\vspace{-0.1in}
\end{table*}%

\subsection{Experimental Results}
We compare the performance of \APN~with 12 prior approaches, namely 
FCLSTM'14 \cite{NIPS2014sequence},
FNN'17 \cite{vaswani2017attention},
DCRNN'17 \cite{li2017diffusion},
TGCN'19 \cite{zhao2019t},
GraphWaveNet'19 \cite{wu2019graph},
STGNN'20 \cite{wang2020traffic},
A3TGCN'21 \cite{bai2021a3t},
STFGNN'21 \cite{li2021spatial},
FRIGATE'22 \cite{gupta2023frigate},
STECAGCN'24 \cite{liu2024spatial},
FCSTGNN'24 \cite{wang2024fully}, and
STGAFormer'24 \cite{geng2024stgaformer}.
%
%
As mentioned earlier, we perform forecasting for the next 6, 12, and 18 measurement periods, which correspond to  30, 60, and 90 minutes for the Abilene and 90, 180, and 270 minutes for the GEANT datasets.
The results are presented in Tables \ref{tab:Abilene} and \ref{tab:GEANT}.
The best results are in bold in the tables.
From these two tables, we make the following observations.

\subsubsection{Overall Observations}
\APN~outperforms all prior methods in almost all settings.
Across the two datasets, our model achieves improvements of 32.14\% in MAE, 28.30\% in RMSE, and 20.47\% in SMAPE compared to the average performance of all graph-based baseline methods across all forecasting lengths.
Specifically, on the Abilene dataset, our model achieves improvements of 44.83\% in MAE, 38.36\% in RMSE, and 23.60\% and on the GEANT dataset, it achieves improvements of 19.45\% in MAE, 18.24\% in RMSE, and 17.33\% in SMAPE.
Looking at the values of the performance metrics for individual approaches in the two tables, while in some cases, some approaches seem to perform slightly better in terms of MAE and RMSE, in terms of SMAPE, \APN~always outperforms all other approaches.
We argue that SMAPE is the most representative metric of the true performance of an approach because it calculates the error in an estimated value relative to the ground truth value.
MAE and RMSE aggregate all errors together, which means that if an approach makes large error in estimating a value that has a small ground truth, that large error may not impact the final MAE and RMSE value if the dataset also has large ground truth values.
SMAPE eliminates the impact of the size of the ground truth values by normalizing them, and thus does not suffer from this limitation of MAE and RMSE.

\subsubsection{Forecasting Horizon}
On the Abilene dataset, \APN's performance stays relatively consistent for the three forecasting horizon's while on the GEANT dataset, \APN's performance experiences slight deterioration with the increasing forecasting horizon.
This is because for the Abilene dataset, the farthest forecast into future is 90 minutes out while for the GEANT dataset, it is 270 minutes out.
Consequently, the impact of error accumulation is more pronounced in the GEANT dataset compared to the Abilene dataset.

\subsubsection{Observations on Prior Approaches}
While \APN~outperforms all prior approaches, it is worthwhile to study how prior approaches compare with each other.
FCLSTM and FNN exhibit relatively poor performance, as they are temporal-only models and cannot capture the hidden relationships between neighboring nodes.
The explicit sequential information modeled by the hidden states gives FCLSTM a greater capacity to handle various temporal patterns compared to FNN, which is limited by its simple architecture.
Graph-based methods, which consider both spatial and temporal dependencies, are more suitable for complex, location-relevant network traffic.
Compared to temporal-only models, most graph-based methods perform better.
FCSTGNN achieves the lowest MAE for 30-minute forecasting on the Abilene dataset, whereas TGCN, which integrates gated recurrent units and graph convolutional networks, demonstrates the best performance under the RMSE metric for the same forecasting horizon.
It is noteworthy that A3TGCN and GraphWaveNet, both graph-based methods, perform relatively poorly in some cases.
This is most likely due to not handling noise in the spatial information.
In \APN, we specifically handle the noise, as described when formulating the spatio-temporal adjacency matrix as well as when performing softmax normalization.
This gives \APN~edge over prior such approaches.
On the GEANT dataset, GraphWaveNet achieves the highest performance under the RMSE metric for 270-minute forecasting, while STGAFormer outperforms other methods in terms of MAE for the same forecasting length.
In contrast, A3TGCN and STGNN fall behind, indicating that simply adding an attention mechanism or a transformer may not yield optimal results due to a tendency to overfit.
%

\subsection{Impact of $p$ in Top-$p$ Percentile}
\label{subsec:Impacetoftop-p}
Recall from Sec. \ref{subsec:Spatio-temporalAdjacencyMatrix} that to keep the spatio-temporal adjacency matrix, $A_{ST}$, from getting too dense, \APN~applies a filter where it sets all entries in $A_{ST}$ that fall in the top-$p$ percentile as one and the remaining as zero.
To investigate the impact of the value of $p$, we evaluate \APN~using various values of $p$, ranging from 20 \% to 80 \%, and forecast the values in the $12$ future measurement periods. 
All other experimental settings remain the same, as described in Sec. \ref{subsec:ExperimentalSettingsandMetrics}.
Figs. \ref{fig:AblationTopPMAE}, \ref{fig:AblationTopPRMSE}, and \ref{fig:AblationTopPSMAPE} show the results.

\begin{figure*}[htb]
  \centering
    \begin{subfigure}{0.325\linewidth}
    \centering
    \includegraphics[width=\linewidth]{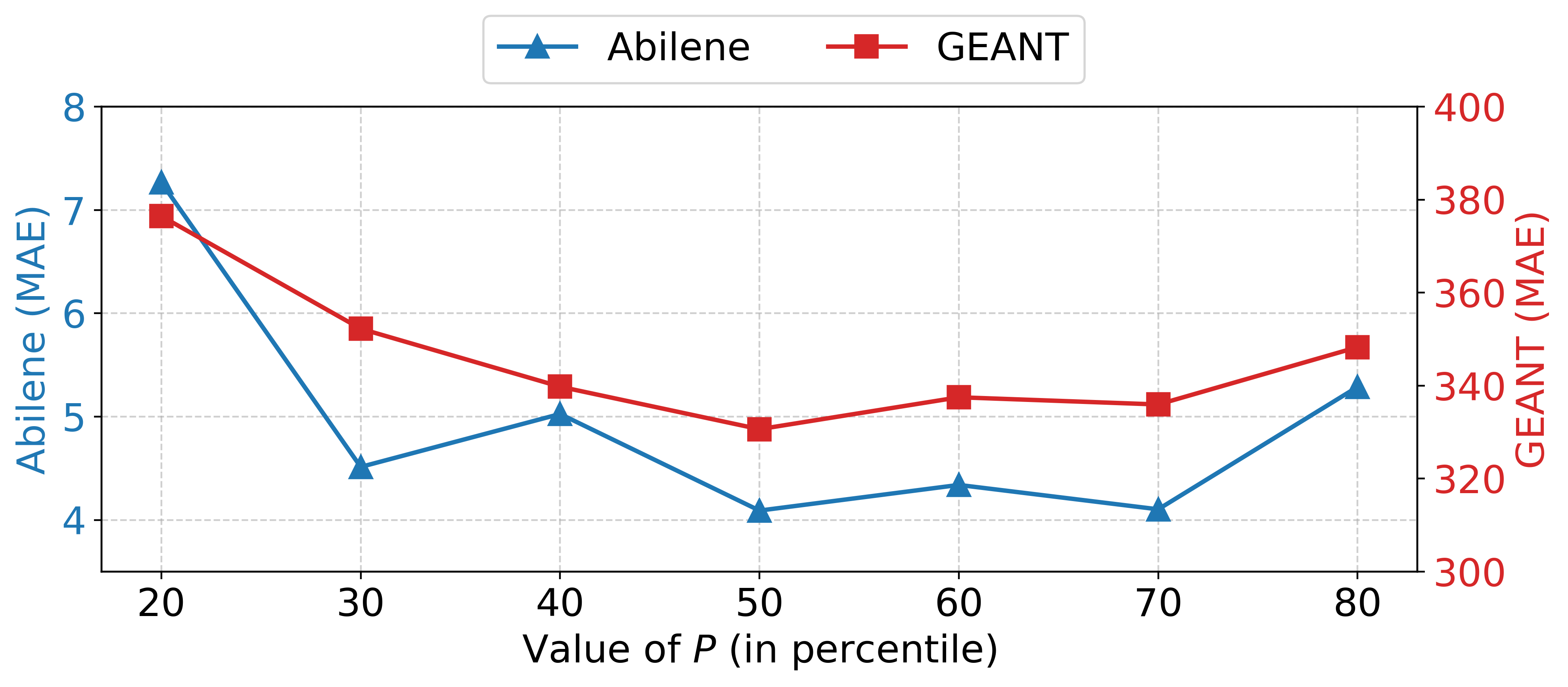}
    \vspace{-0.2in}
    \caption{MAE}
    \label{fig:AblationTopPMAE}
  \end{subfigure}
  \hspace{-0.05in}
  \begin{subfigure}{0.325\linewidth}
    \centering
    \includegraphics[width=\linewidth]{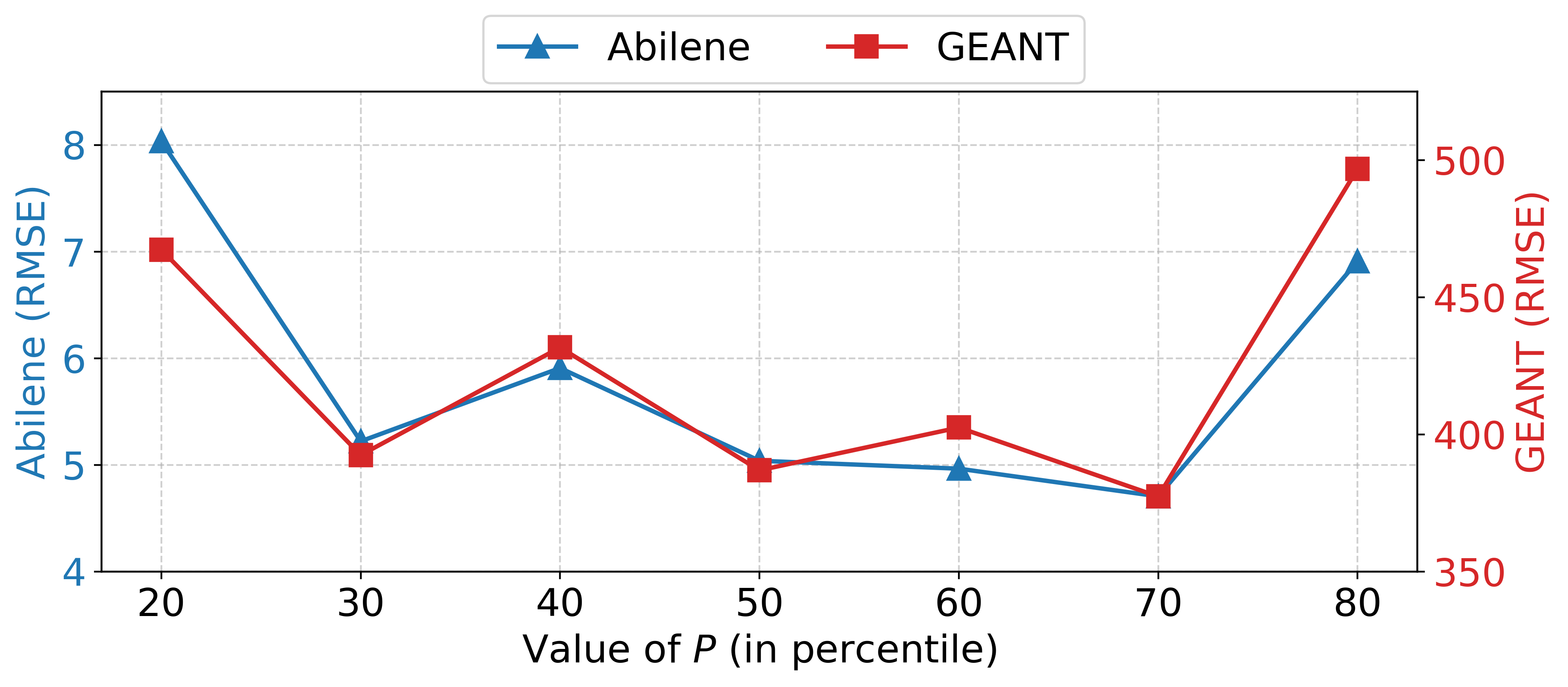}
    \vspace{-0.2in}
    \caption{RMSE}
    \label{fig:AblationTopPRMSE}
  \end{subfigure}
  \hspace{-0.05in}
  \begin{subfigure}{0.325\linewidth}
    \centering
    \includegraphics[width=\linewidth]{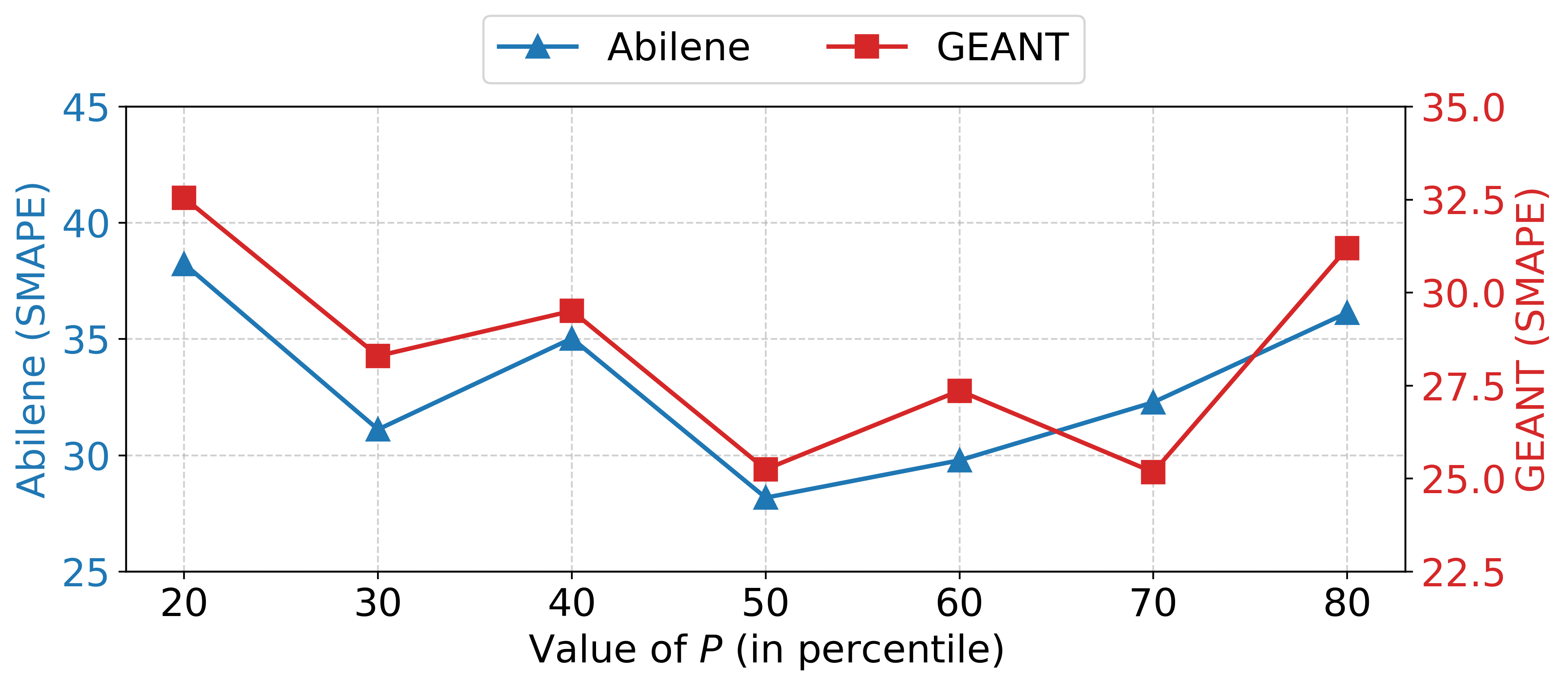}
    \vspace{-0.2in}
    \caption{SMAPE}
    \label{fig:AblationTopPSMAPE}
  \end{subfigure}
    \vspace{-0.05in}
  \caption{Impact of the value of $p$ when selecting top-$p$ percentile values in $A_{ST}$.}
    \vspace{-0.2in}
  \label{fig:Ablation topP}
\end{figure*}

We observe from these figures that NetSight's performance on both datasets, across all three metrics, follows a U-shaped trend.
The error is high at smaller values of $p$ due to significant loss of spatio-temporal information and is also high at larger values of $p$ due to allowing too much noise to stay in $A_{ST}$.
The best overall performance occurred at $p = 50\%$ in terms of MAE and SMAPE, and at $p = 70\%$ in terms of RMSE.
$p = 50\%$ still yielded the second-best performance in RMSE. 
Notably, RMSE increases sharply beyond $p = 70\%$, indicating that the additional noise introduced at such high values of $p$ exceeds the robustness of \APN.
This degradation could potentially be mitigated by employing a larger model and/or stronger regularization.
The necessity of a properly constructed spatio-temporal matrix, $A_{ST}$, is evident as even the lowest performance of \APN seen in Figs. \ref{fig:AblationTopPMAE}, \ref{fig:AblationTopPRMSE}, and \ref{fig:AblationTopPSMAPE} surpasses nearly half of the state-of-the-art prior approaches.

Based on the finding that NetSight's error is a convex function of $p$, a practical strategy to select the value of $p$ would be to initialize a $p_\text{L}=0$ and a $p_\text{R}=100$ and set $(p=p_\text{L}+p_\text{R})/2$.
If the error at $(p+p_\text{R})/2$ is lower than the error at $(p_\text{L}+p)/2$, then set $p_\text{L}=p$.
Otherwise, set $p_\text{R}=p$.
Repeat until the difference between the errors at $(p+p_\text{R})/2$ and $(p_\text{L}+p)/2$ is less than a threshold.
To ensure that the selected value of $p$ is not a noise-induced erroneous minima, measure the errors at $p\pm\epsilon$ for a small pre-selected value of $\epsilon$.
If the difference between the errors at $p\pm\epsilon$ is still below the threshold, then this value of $p$ is the optimal to use.
Otherwise, add random values (between 0 and 50) to the previous values of $p_\text{L}$ and $p_\text{R}$ and repeat the steps.
This approach is, in essence, a mixture of gradient descent and binary search algorithms.

\vspace{-0.01in}
\subsection{Ablation Study}
\vspace{-0.01in}
Next, we study the importance of various components that constitute \APN~through ablation experiments.
%

\vspace{-0.01in}
\subsubsection{Impact of Spatial and Temporal Modeling}
To study the effectiveness of modeling spatial dependencies and temporal dependencies, we create two variants of \APN.
%
\begin{enumerate}[leftmargin=*]
\item $\APN_{\text{$\backslash$Sp}}$, where the local (Sec. \ref{subsubsec:LocalSpatialDependencyModeling}) and global spatial modeling (Sec. \ref{subsubsec:GlobalSpatialDependencyModeling}) is not performed, leaving only the transformer and the multi-head prediction layers. This configuration captures only the temporal dependencies.
\item $\APN_{\text{$\backslash$Tm}}$, where the global and local temporal modeling (Sec. \ref{subsec:Temporal Dependency Modeling}) is not performed. The outputs of the node normalization layer and the pooling process are combined and fed into the prediction layer. This configuration captures only the spatial dependencies.
\end{enumerate}

The results are illustrated in Fig. \ref{fig:Ablation Results}.
All results in this section are for predicting 12 future time steps.
We can clearly see from Fig. \ref{fig:Ablation Results} that both $\APN_{\text{$\backslash$Sp}}$ and $\APN_{\text{$\backslash$Tm}}$ perform poorly compared to \APN, highlighting the necessity of modeling both spatial and temporal dependencies.
$\APN_{\text{$\backslash$Tm}}$ outperforms $\APN_{\text{$\backslash$Sp}}$ in most cases, suggesting that the dataset contains more spatial dependencies compared to temporal dependencies.
Although neither $\APN_{\text{$\backslash$Tm}}$ nor $\APN_{\text{$\backslash$Sp}}$ surpass \APN, they still outperform several of the baseline methods in Tables \ref{tab:Abilene} and \ref{tab:GEANT}, demonstrating the effectiveness of our proposed method.

\begin{figure}[htbp]
  \centering
 \vspace{-0.1in}
  \begin{subfigure}{0.49\linewidth}
    \centering
    \includegraphics[width=\linewidth]{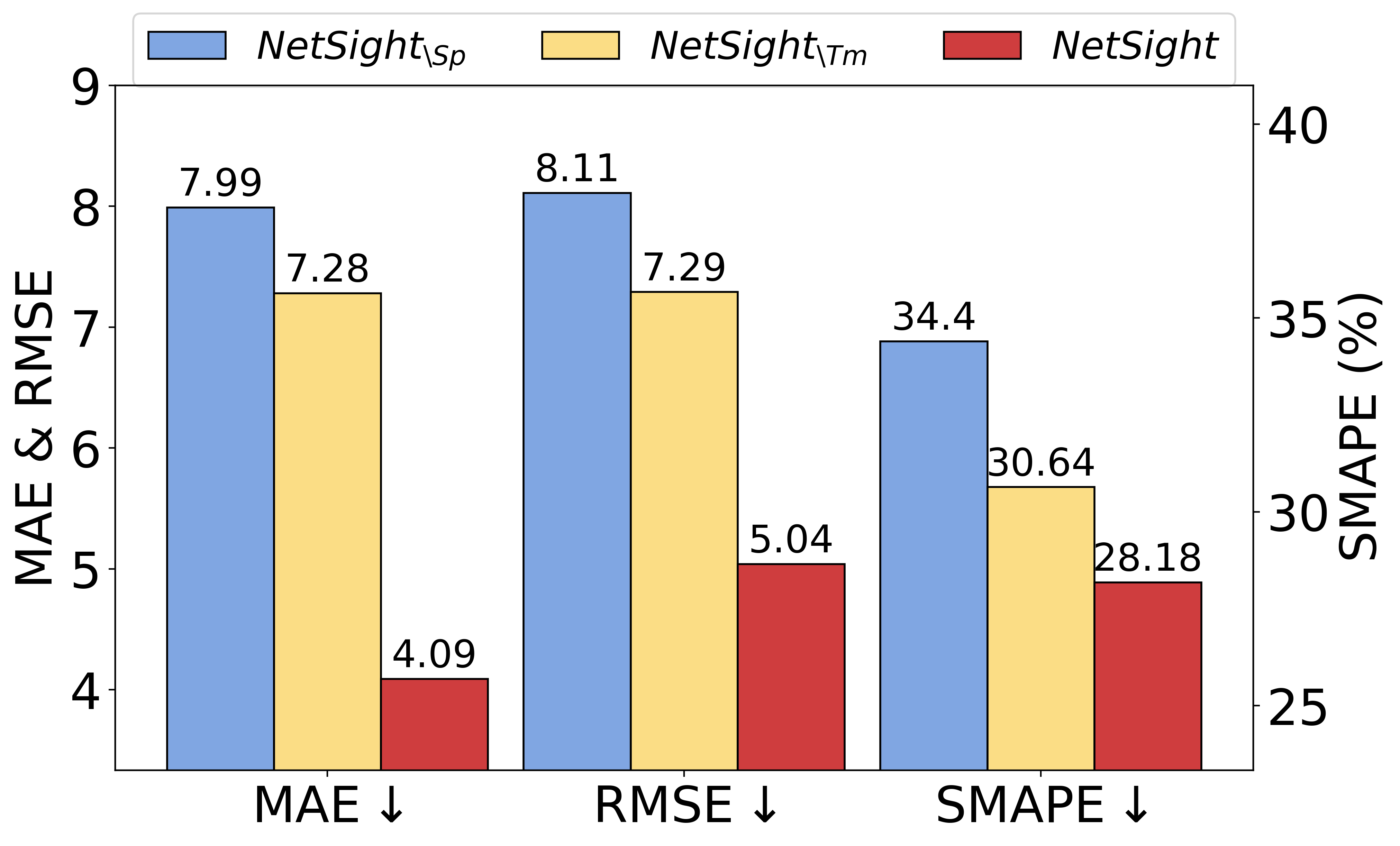}
    \vspace{-0.2in}
    \caption{Abilene Dataset}
    \label{Ablation Abilene}
  \end{subfigure}
  \hfill
  \begin{subfigure}{0.49\linewidth}
    \centering
    \includegraphics[width=\linewidth]{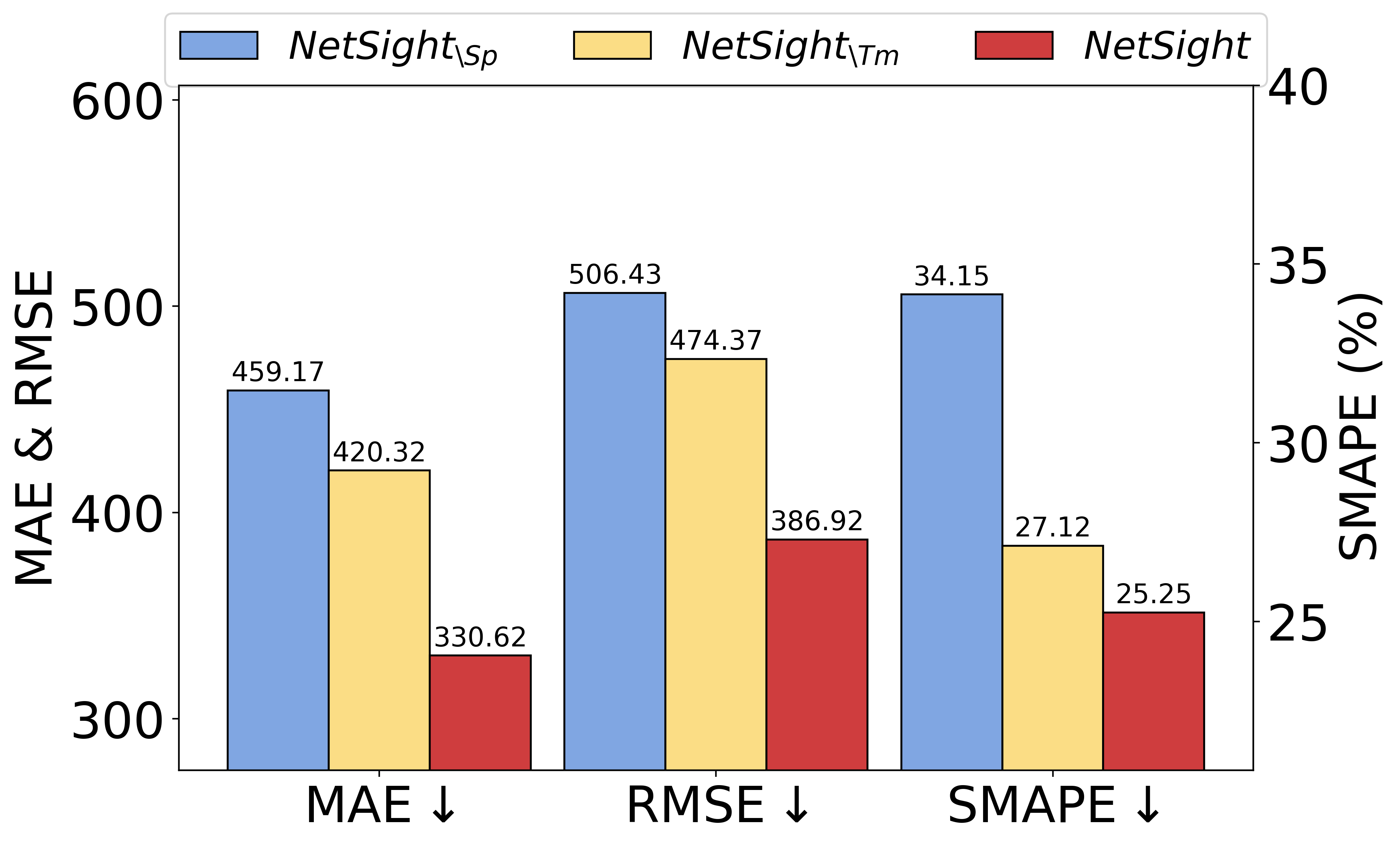}
    \vspace{-0.2in}
    \caption{GEANT Dataset}
    \label{Ablation GEANT}
  \end{subfigure}
    \vspace{-0.2in}
  \caption{Impact of spatial and temporal modeling blocks.}
    \vspace{-0.075in}
  \label{fig:Ablation Results}
\end{figure}

\vspace{-0.01in}
\subsubsection{Impact of Node Normalization}
To study the effectiveness of the node normalization approach proposed in Sec. \ref{subsubsec:LocalSpatialDependencyModeling}, we create another variant of \APN, namely $\APN_{\text{$\backslash$NN}}$, which includes everything in \APN~except this node normalization procedure.
Fig. \ref{fig:Ablation huberLoss} shows the Huber loss during the first 50 training epochs, with and without node normalization.
It is evident that node normalization significantly speeds up the convergence process during training by more rapidly reducing the loss function.
While we observed a few sudden peaks in the GEANT dataset around the 20th epoch, the loss quickly stabilized within a few epochs.
%

\begin{figure}[htbp]
  \centering
    \vspace{-0.1in}
    \begin{subfigure}{0.49\linewidth}
    \centering
    \includegraphics[width=\linewidth]{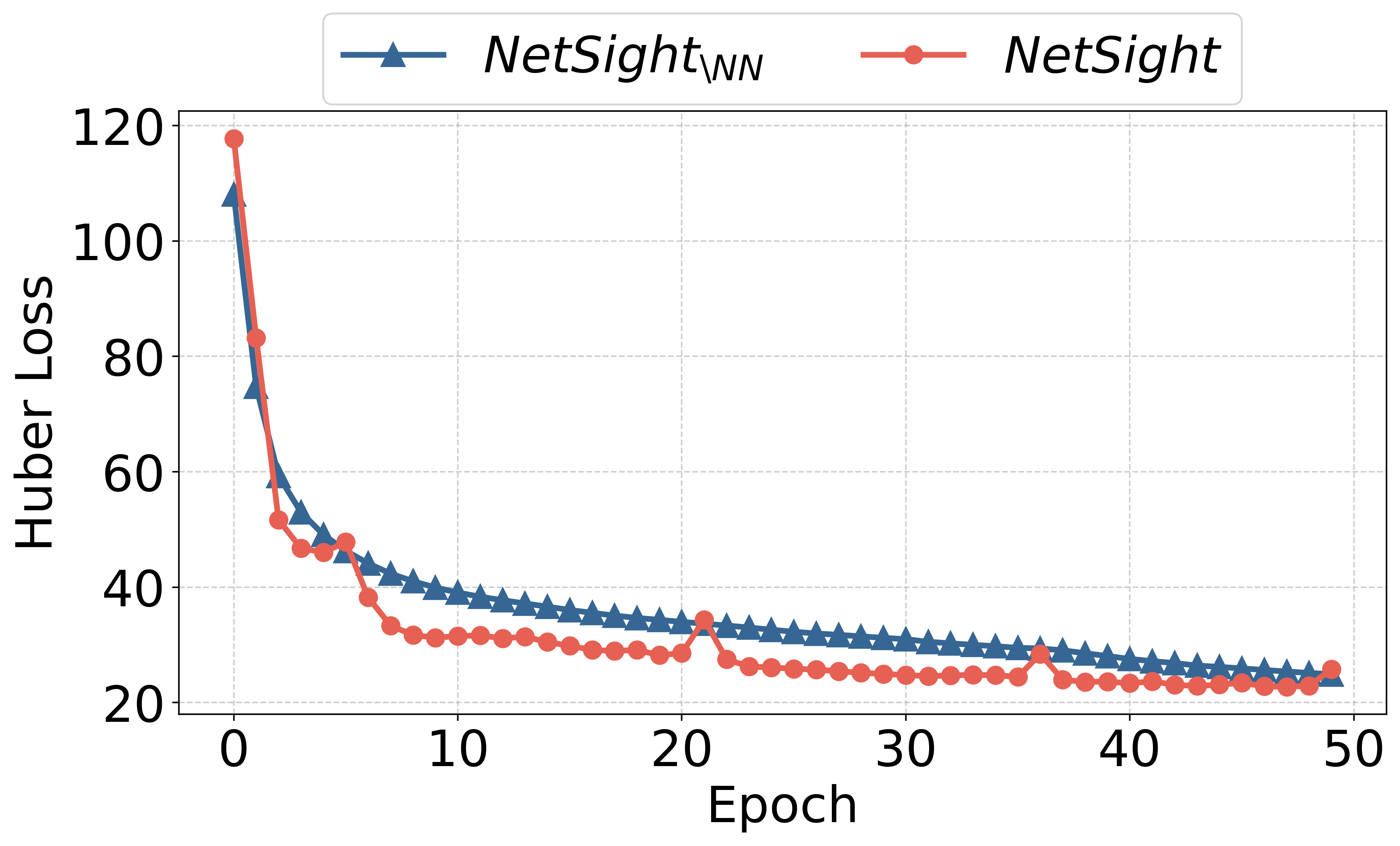}
    \vspace{-0.2in}
    \caption{Abilene Dataset}
    \label{Ablation Abilene huberLoss}
  \end{subfigure}
  \hfill
  \begin{subfigure}{0.49\linewidth}
    \centering
    \includegraphics[width=\linewidth]{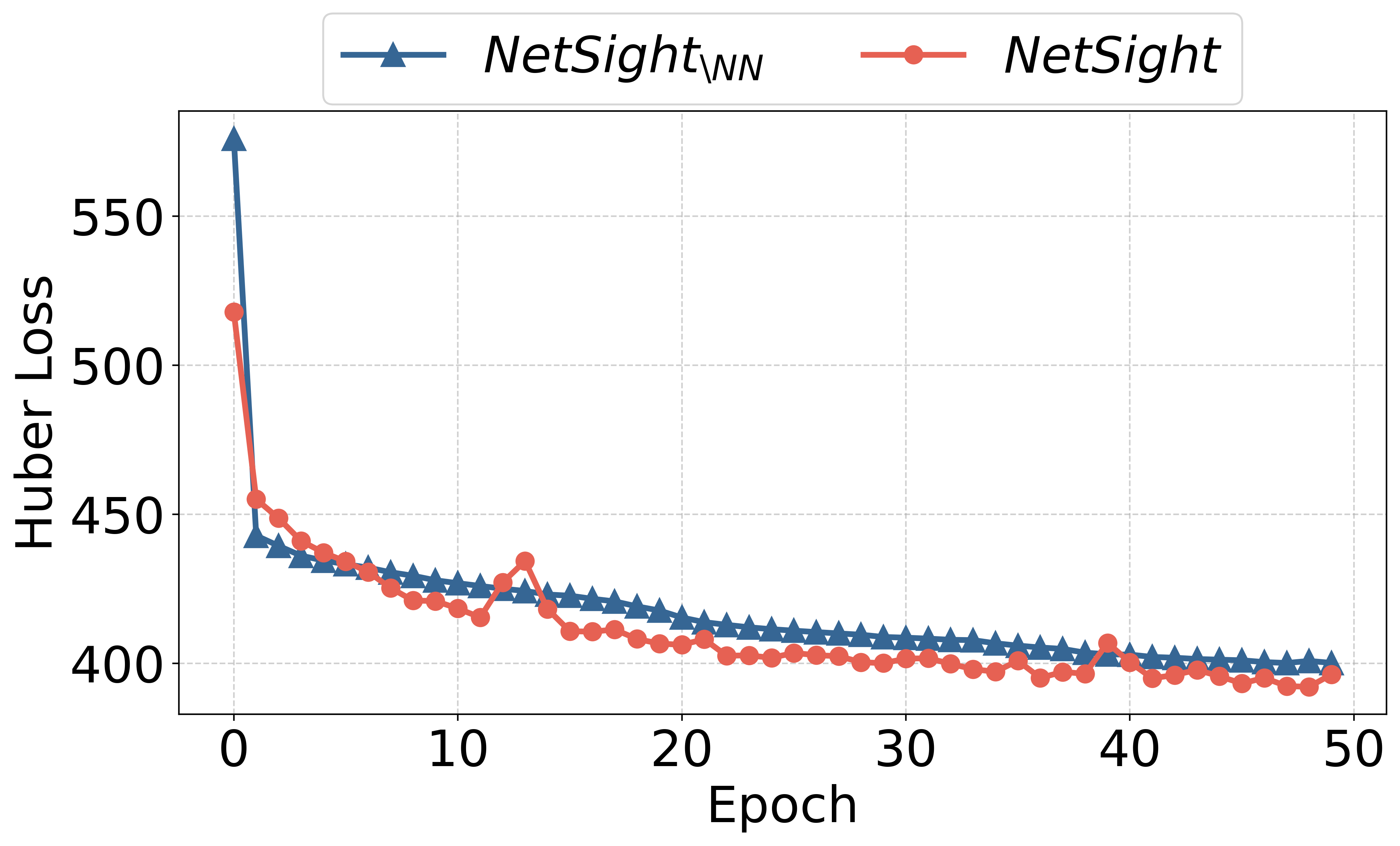}
    \vspace{-0.2in}
    \caption{GEANT Dataset}
    \label{Ablation GEANT huberLoss}
  \end{subfigure}
    \vspace{-0.2in}
  \caption{Impact of node normalization in spatial modeling.}
    \vspace{-0.075in}
  \label{fig:Ablation huberLoss}
\end{figure}

\vspace{-0.01in}
\subsubsection{Impact of Salient Components of \APN}
We investigated the performance of \APN~without three of its other important components: 
\begin{enumerate}[leftmargin=*]
\item Without spatio-temporal adjacency matrix (\ie, using only $A_S$ and not $A_{ST}$).
\item Without the pooling mechanism, which captures global spatial information, that we described in Sec. \ref{subsubsec:GlobalSpatialDependencyModeling}.
\item Without the attention based prediction layer and using a simple multi-layer perceptron (MLP) instead.
\end{enumerate}
Table \ref{tab:ablationAbGE} shows the results.
We again see that the performance of \APN~deteriorates without each of these components, with the spatio-temporal adjacency matrix demonstrating the highest importance, followed by the pooling mechanism, and then the attention based prediction layer.
The spatio-temporal fusion matrix effectively captures hidden local temporal dependencies, while the pooling mechanism facilitates a comprehensive aggregation of neighboring information to generate a global contextual understanding.
The multi-head attention mechanism enables the prediction layer to sequentially predict and thus outperforms MLP.
Notably, although none of these variants of \APN~surpass the complete \APN~framework, each variant still outperforms several prior approaches, substantiating the effectiveness of our proposed approach.

\begin{table}[htbp]
\vspace{-0.1in}
\centering
  \caption{Impact of various components of \APN.}
\vspace{-0.05in}
  \begin{tabular}{l|ccc|ccc}
    \toprule
                    &\multicolumn{3}{c|}{Abilene}                   &\multicolumn{3}{c}{GEANT}\\
    \cline{1-7}
    Without         & MAE           & RMSE          & SMAPE        & MAE             & RMSE          & SMAPE\\
    \cline{2-7}
    $\drsh$$A_{ST}$    & 6.54          & 6.69          & 30.5         & 417.9          & 425.6          & 27.1\\
    $\drsh$Pooling     & 6.02          & 6.17          & 29.7         & 377.3          & 417.7          & 26.4\\
    $\drsh$Att. Pred.  & 5.19          & 5.80          & 29.2         & 351.1          & 397.3          & 25.7\\
    \midrule
    \APN            & \textbf{4.09} & \textbf{5.04} & \textbf{28.2}& \textbf{330.6} & \textbf{386.9} & \textbf{25.3} \\
    \bottomrule
  \end{tabular}
  \label{tab:ablationAbGE}
\vspace{-0.1in}
\end{table}

\section{Conclusion}
In this paper, we proposed \APN, the first framework that models joint spatio-temporal dependencies simultaneously at both global as well as local scales to predict future traffic measurements.
The key technical novelties of \APN~lie in the design of the adjacency matrix that quantifies both spatial and temporal proximity among nodes, in the design of the node normalization approach to mitigate covariance shifts, and in the adoption of an attention mechanism to increase the prediction horizon.
The key technical depth of \APN~lies in its ability to simultaneously capture global and local joint spatio-temporal dependencies, which has not been accomplished previously.
We implemented and extensively compared \APN~with 12 state-of-the-art approaches.
The results show that for the Abilene dataset, \APN~achieves improvements of up to 44.83\% in MAE, 38.36\% in RMSE, and 23.60\% in SMAPE, and for the GEANT dataset, it achieves improvements of 19.45\% in MAE, 18.24\% in RMSE, and 17.33\% in SMAPE compared to the average performance of all graph-based baseline methods across all forecasting lengths.
We also carried out an extensive ablation study, which showed that all components included in the design of \APN~play significant role in helping it achieve its high accuracy.
%

}

\newpage

\bibliographystyle{unsrt}
\bibliography{citation}

\newpage
\clearpage
\newpage

\appendix

\section{Case Study}
\begin{figure}[htbp]
  \centering
  \begin{subfigure}{0.49\linewidth}
    \centering
    \includegraphics[width=\linewidth]{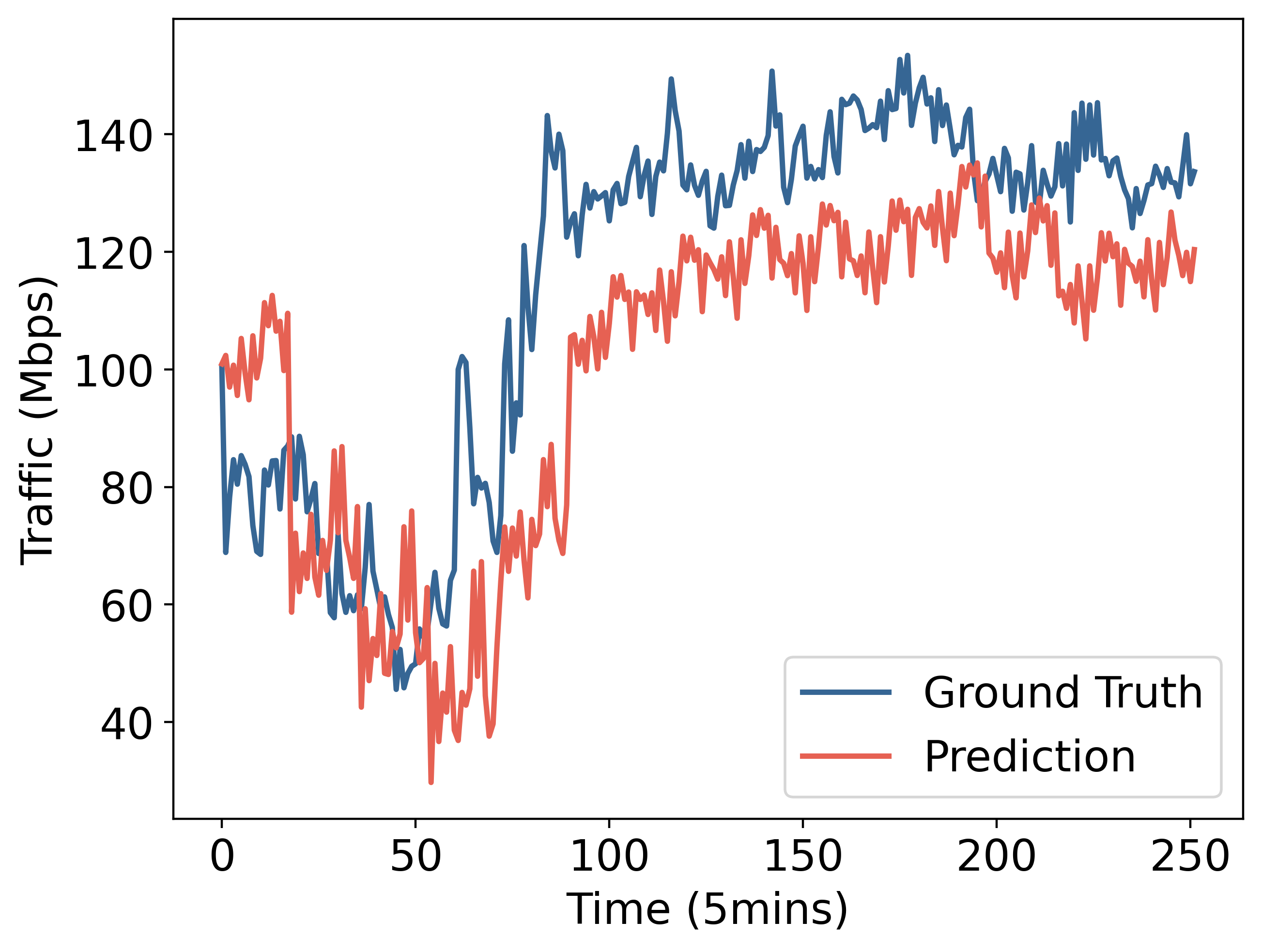}
    \caption{1 Epoch}
    \label{1-epoch-abilene}
  \end{subfigure}
  \hfill
  \begin{subfigure}{0.49\linewidth}
    \centering
    \includegraphics[width=\linewidth]{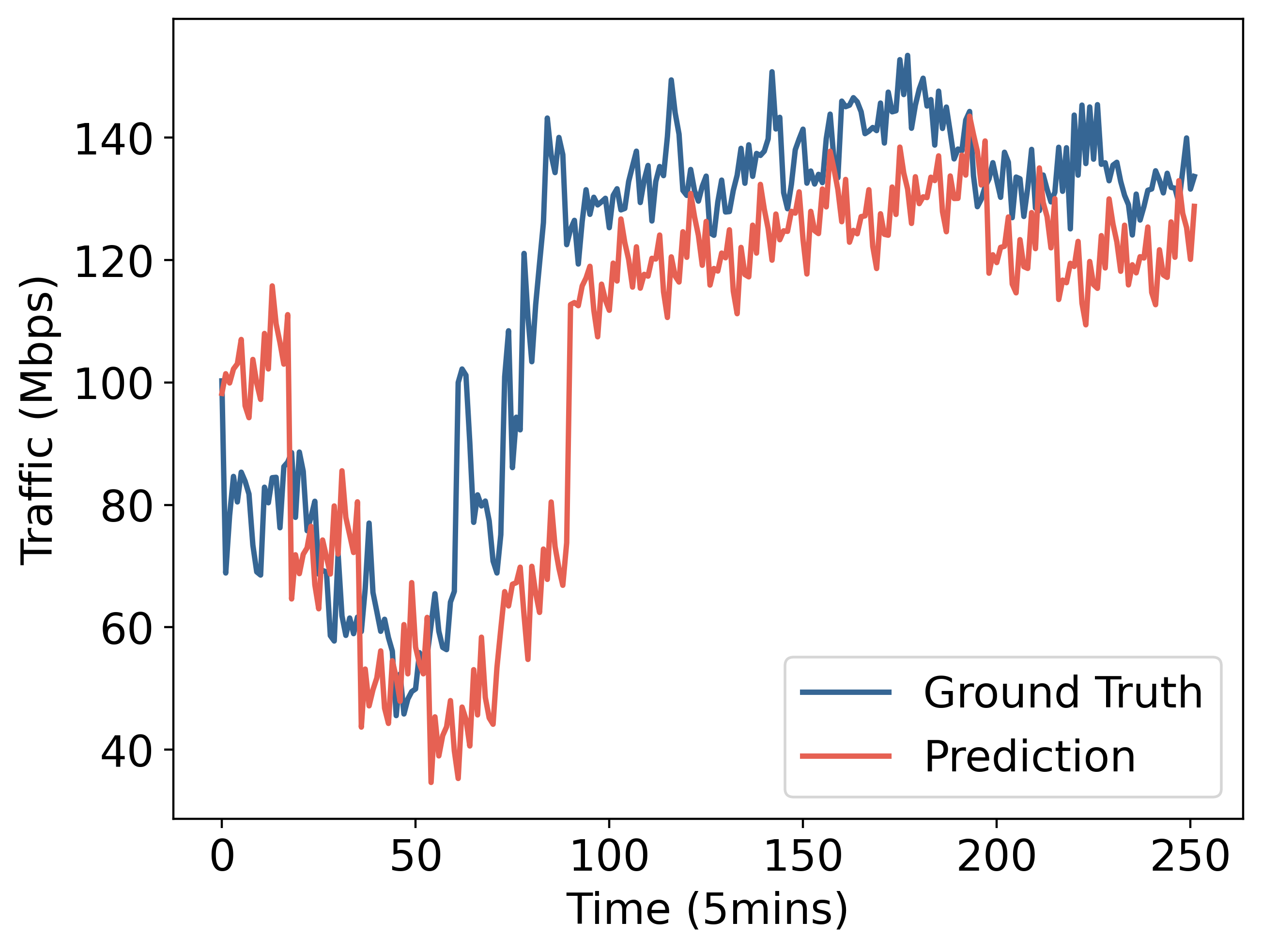}
    \caption{5 Epochs}
    \label{5-epoch-abilene}
  \end{subfigure}
  \vfill
  \begin{subfigure}{0.49\linewidth}
    \centering
    \includegraphics[width=\linewidth]{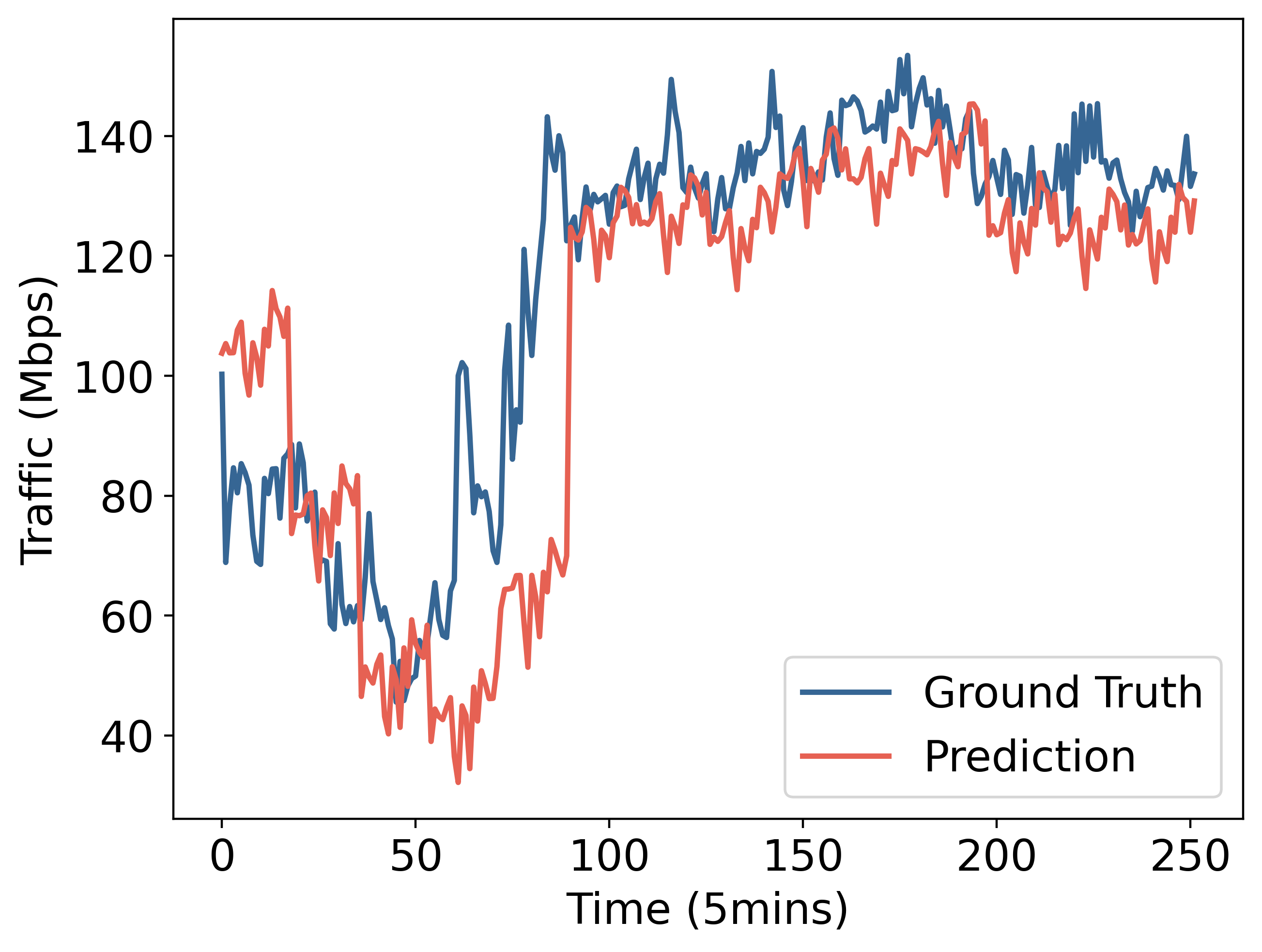}
    \caption{10 Epochs}
    \label{10-epoch-abilene}
  \end{subfigure}
  \begin{subfigure}{0.49\linewidth}
    \centering
    \includegraphics[width=\linewidth]{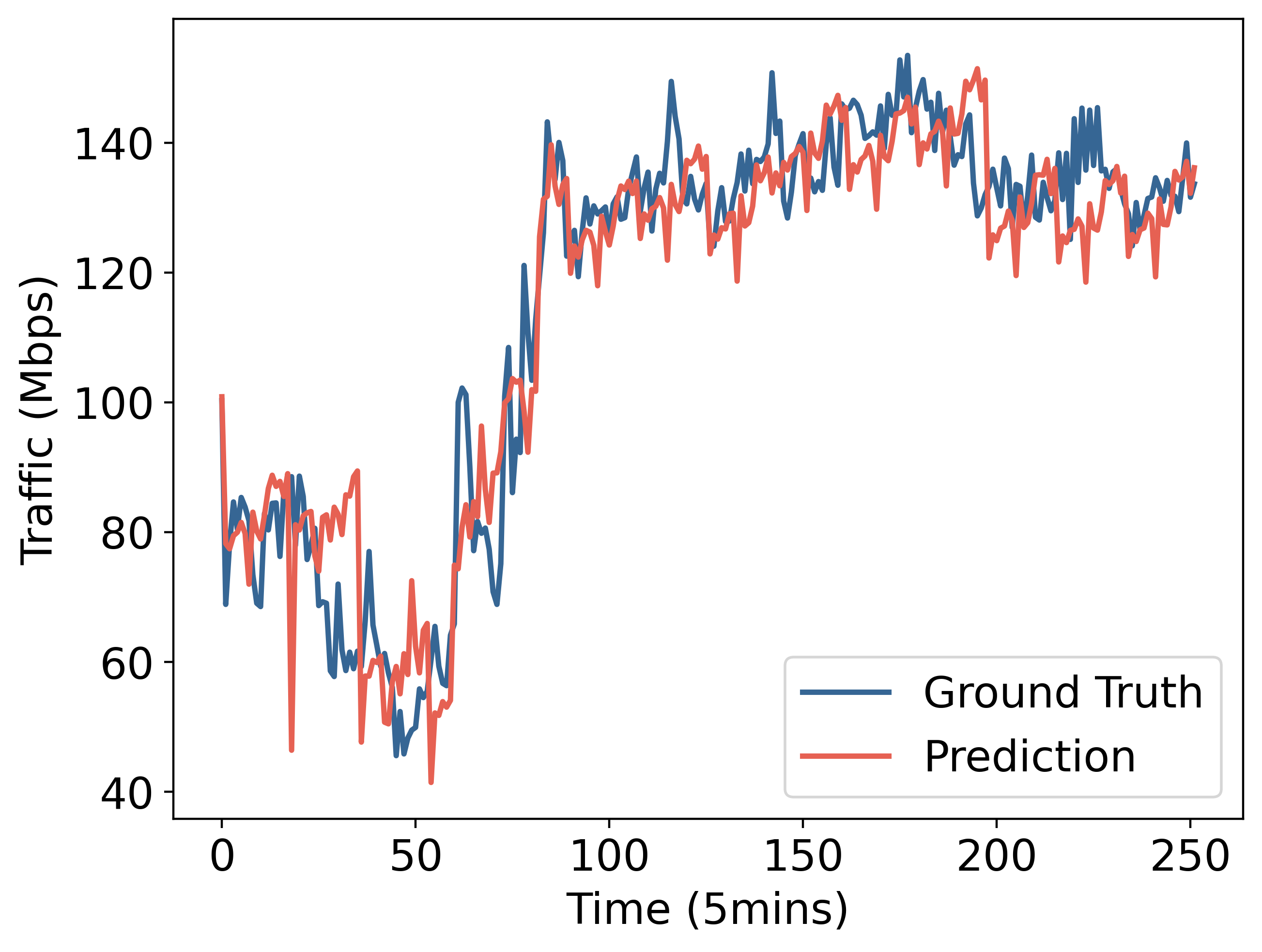}
    \caption{100 Epochs}
    \label{100-epoch-abilene}
  \end{subfigure}
  \caption{Visualization of Forecasting on Abilene}
  \label{fig:casestudy-visualization-abilene}
\end{figure}

\begin{figure}[htbp]
  \centering
  \begin{subfigure}{0.49\linewidth}
    \centering
    \includegraphics[width=\linewidth]{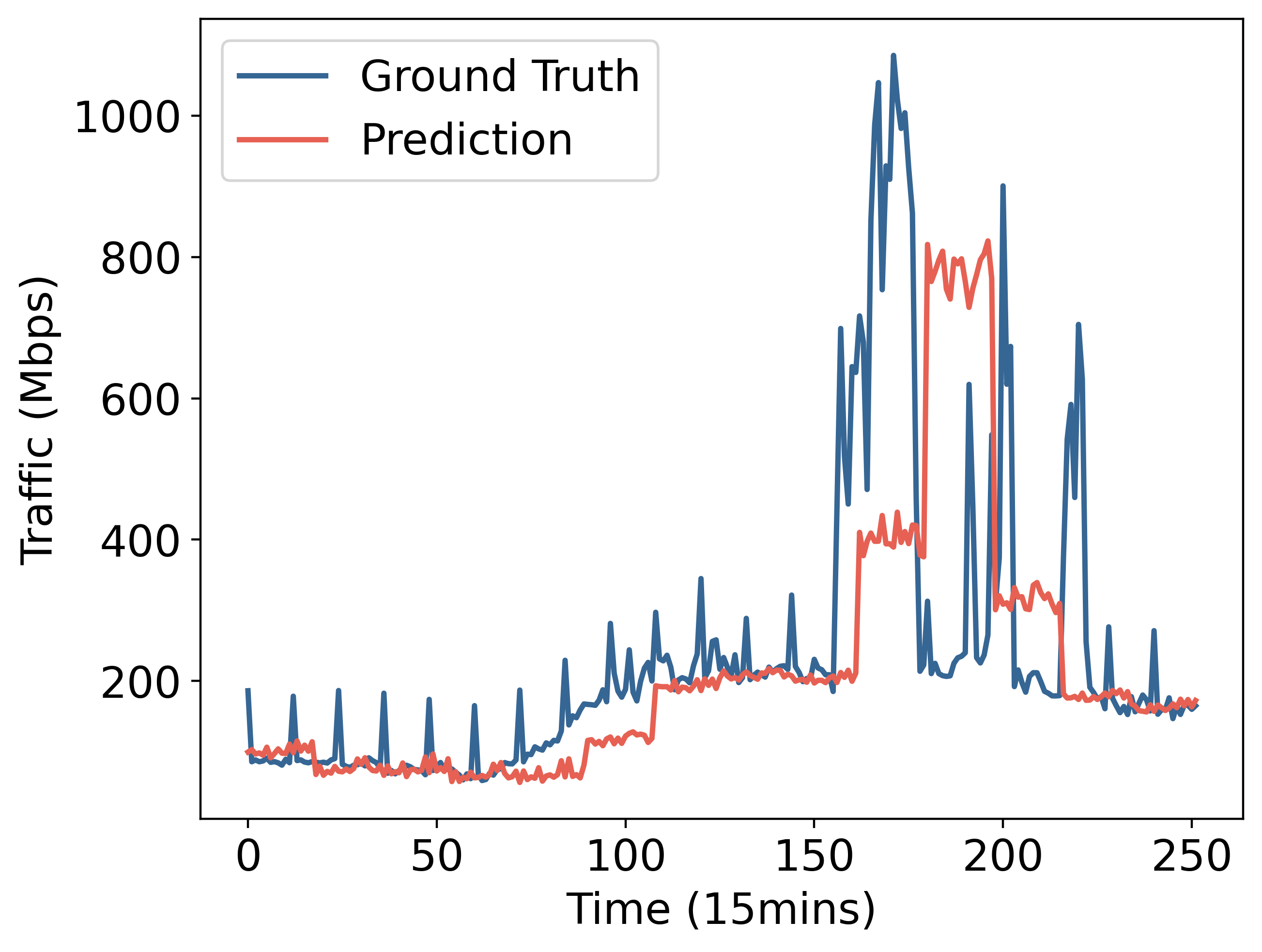}
    \caption{1 Epoch}
    \label{1-epoch-geant}
  \end{subfigure}
  \hfill
  \begin{subfigure}{0.49\linewidth}
    \centering
    \includegraphics[width=\linewidth]{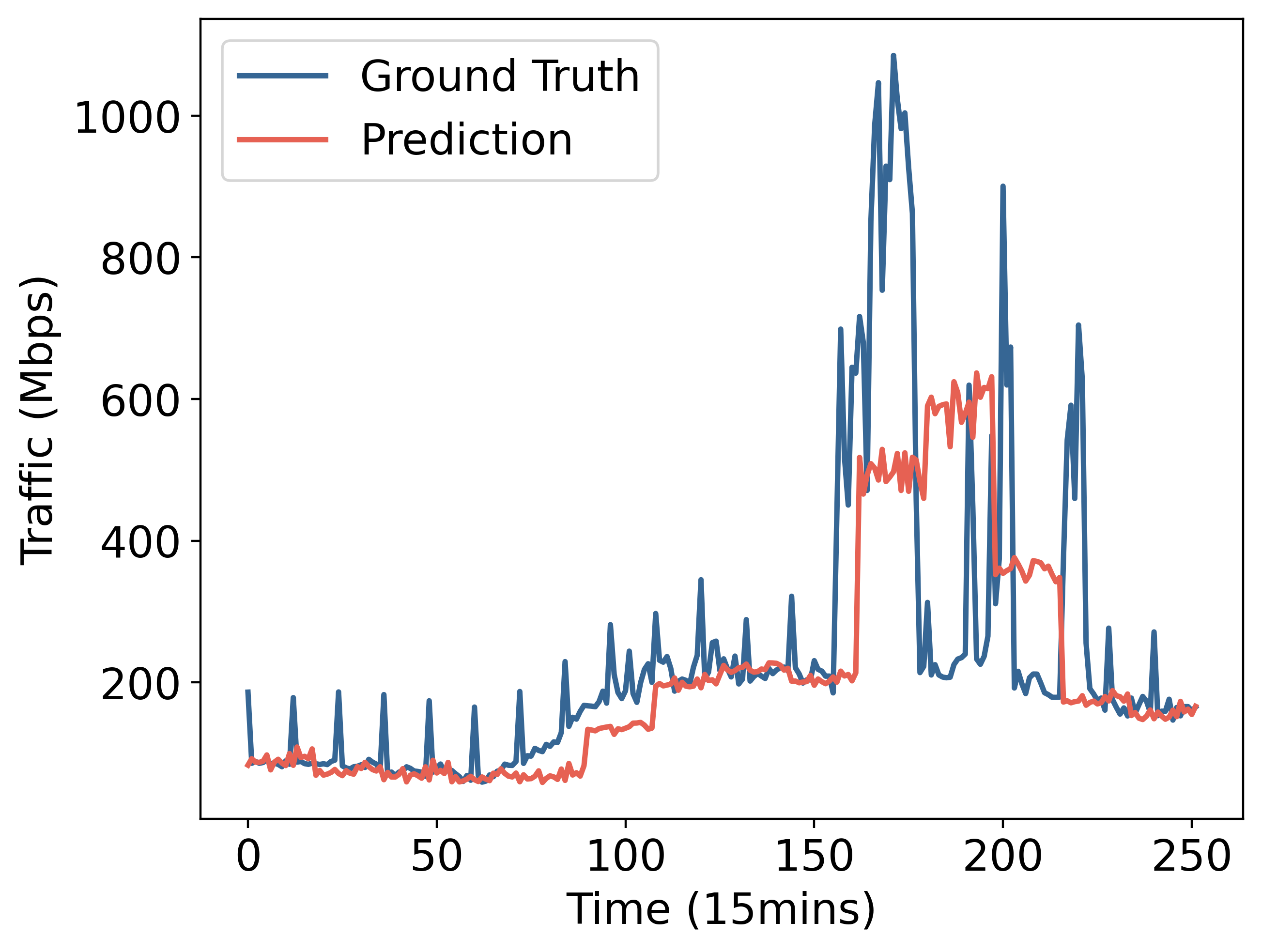}
    \caption{5 Epochs}
    \label{5-epoch-geant}
  \end{subfigure}
  \vfill
  \begin{subfigure}{0.49\linewidth}
    \centering
    \includegraphics[width=\linewidth]{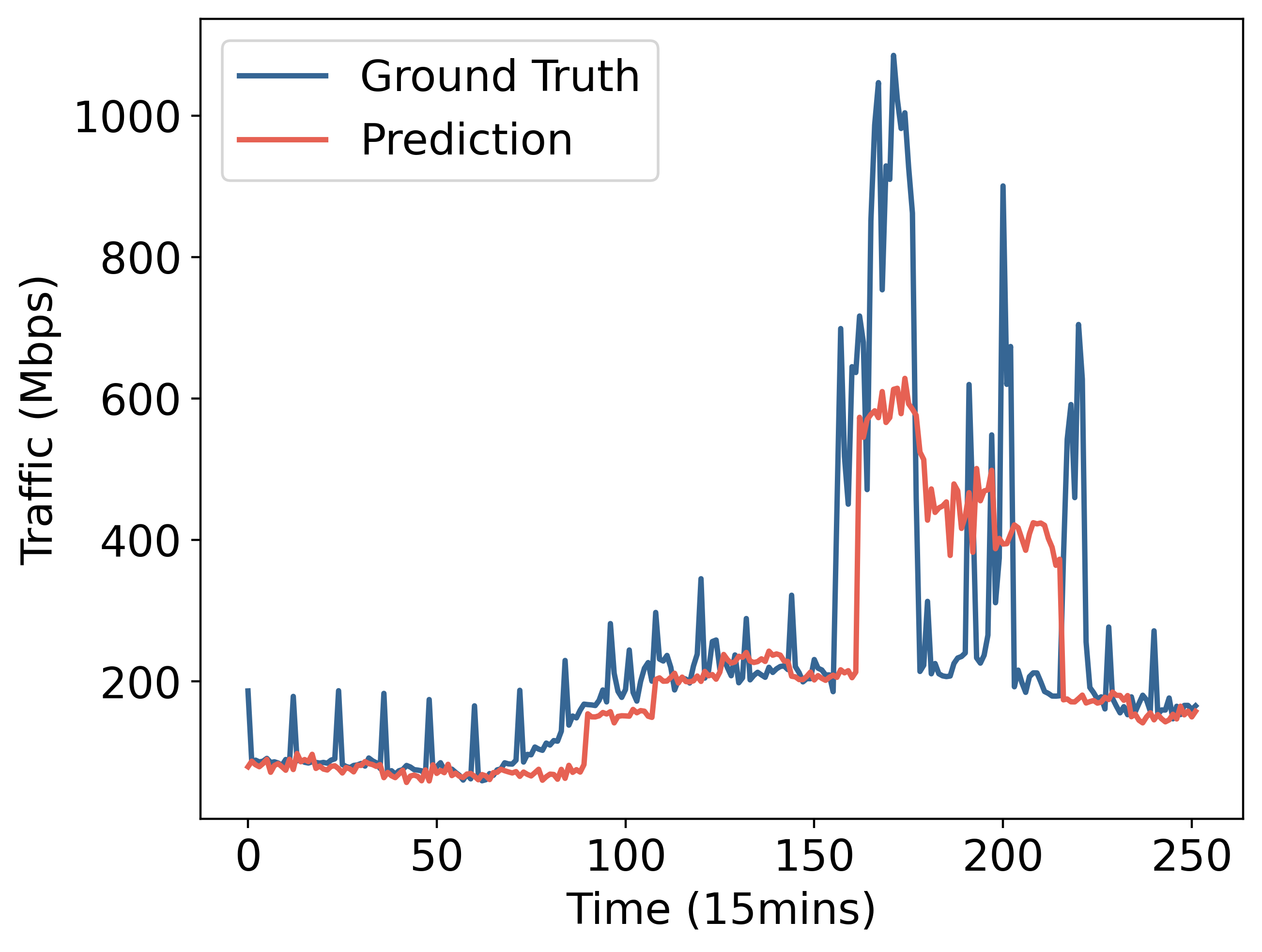}
    \caption{10 Epochs}
    \label{10-epoch-geant}
  \end{subfigure}
  \begin{subfigure}{0.49\linewidth}
    \centering
    \includegraphics[width=\linewidth]{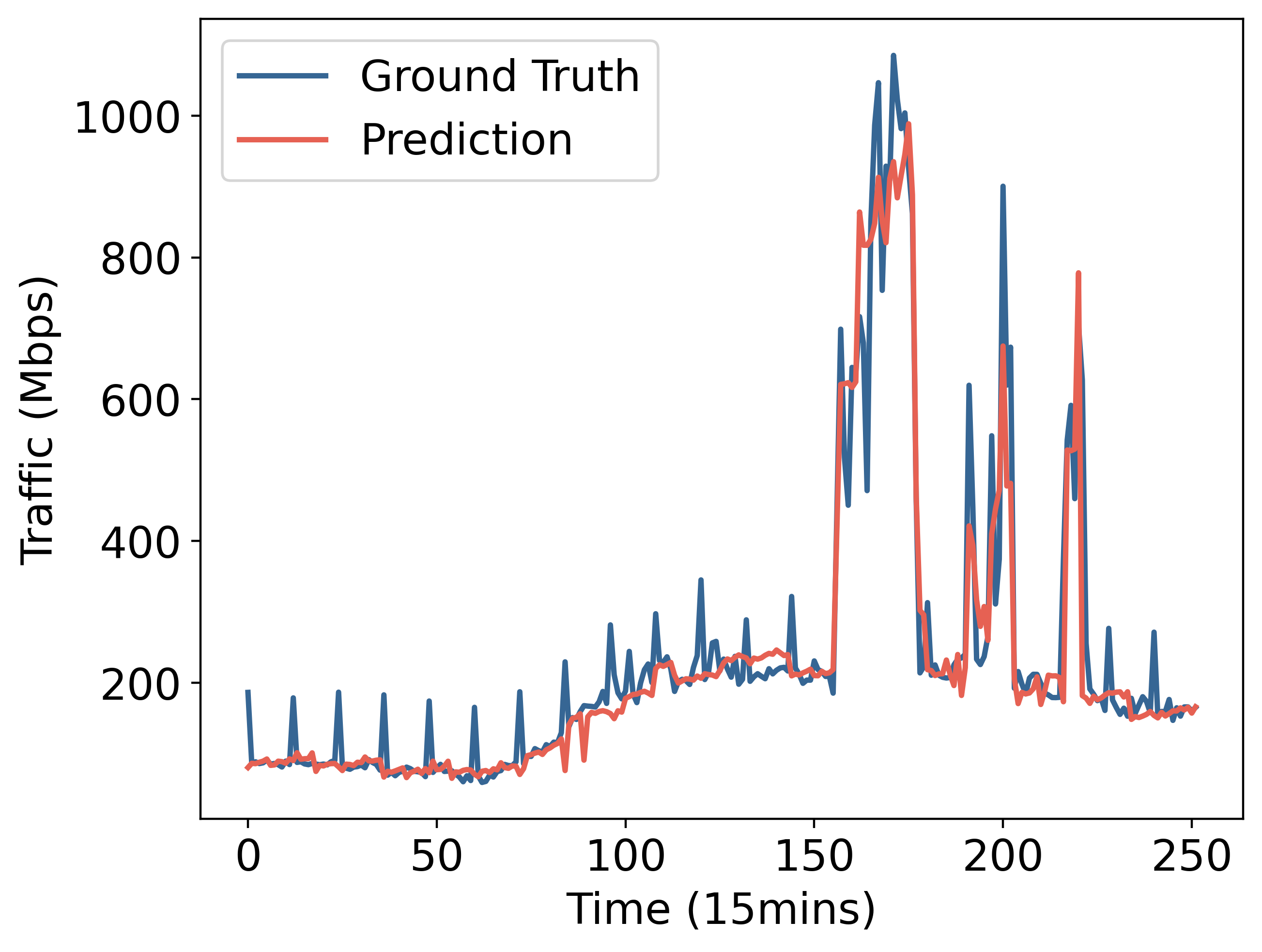}
    \caption{100 Epochs}
    \label{100-epoch-geant}
  \end{subfigure}
  \caption{Visualization of Forecasting on GEANT}
  \label{fig:casestudy-visualization-geant}
\end{figure}

To explore how the model handles complex temporal patterns, we conduct a case study in which we randomly select one node from each dataset and visualize the forecasting curve across different training epochs. We follow the settings in \cite{wang2020traffic}, with the forecasting length set to 18 steps. The results are depicted in Figures \ref{fig:casestudy-visualization-abilene} and \ref{fig:casestudy-visualization-geant}.

On the Abilene dataset, the model's performance improves as the number of training epochs increases. Specifically, even after just one epoch of training, the model is able to capture the overall trend. During the first 10 epochs, as shown in Figures \ref{1-epoch-abilene}-\ref{10-epoch-abilene}, it appears that the model is primarily adjusting its bias parameters, as the overall trend remains largely unchanged while the prediction shifts vertically from 110 Mbps to 130 Mbps, particularly for the time period between 100 and 250. In Figure \ref{100-epoch-abilene}, after training for 100 epochs, the model is able to capture both the overall trend and the finer details more effectively than before. However, there are still several abrupt changes and peaks that the model overlooks. This is likely because the model needs to minimize the loss across all nodes, which typically requires longer training steps and a balanced trade-off between overfitting and generalizability.

A similar observation can be made for the GEANT dataset. The model's performance continues to improve as training progresses. Notably, the traffic in this dataset changes much more drastically compared to the Abilene dataset, with many sudden peaks present in the ground truth. However, after training for 100 epochs, the model is able to successfully handle these drastic changes, particularly from Time 150 to 230. It is worth noting that the model omits some of the temporary sudden peaks like from Time 0 to 150, instead focusing on fitting the overall trend. We do not consider this a drawback, as it prevents overfitting and allows the model to be more generalizable and robust to other temporal patterns. These observations provide detailed insights into how the model works and further demonstrate the effectiveness of our framework.
\section{Congestion Prediction}
\begin{figure}[htbp]
  \centering
  \begin{subfigure}{0.49\linewidth}
    \centering
    \includegraphics[width=\linewidth]{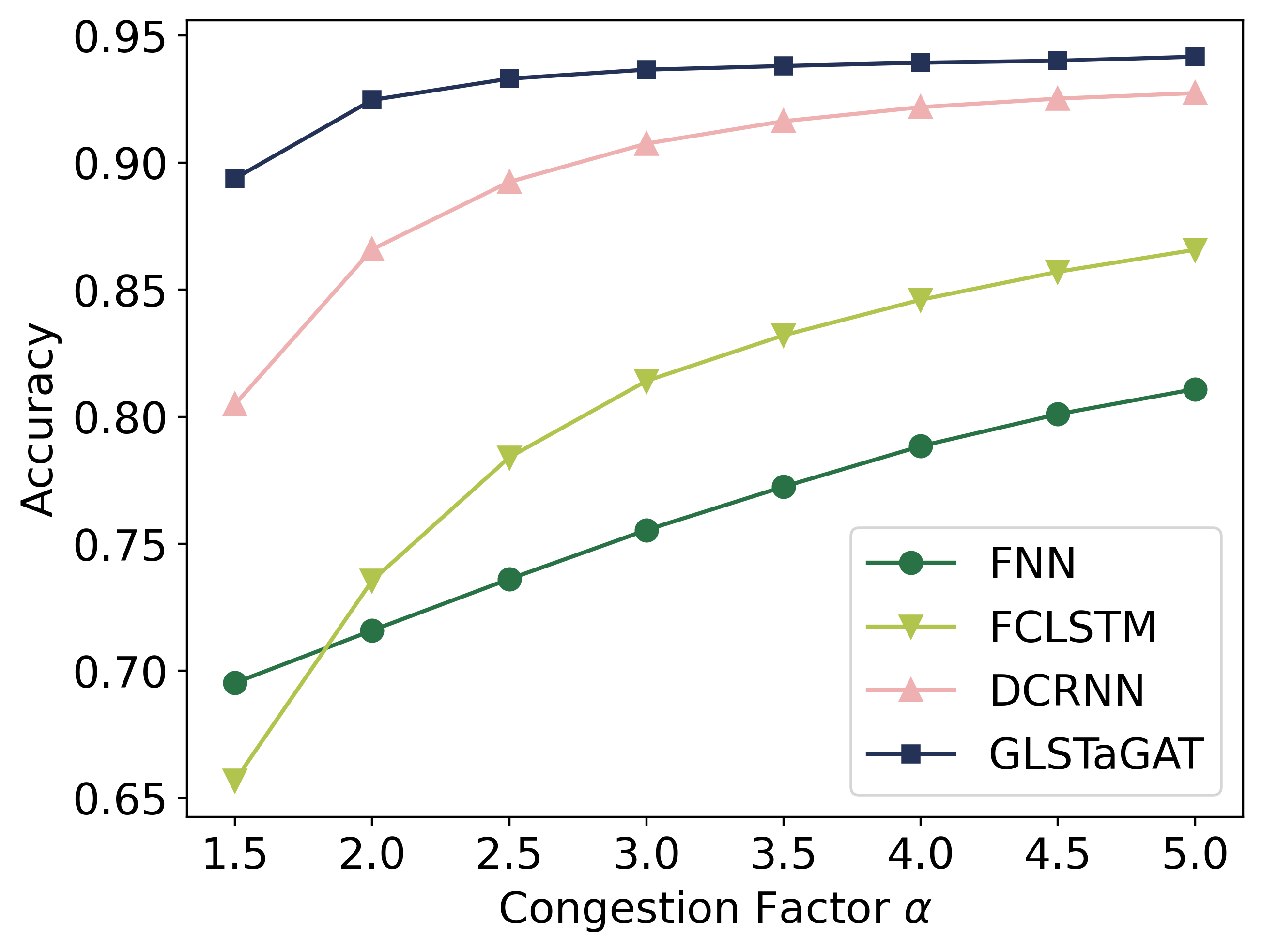}
    \caption{Comparison with Baselines (Part 1)}
    \label{accu-abilene-1}
  \end{subfigure}
  \hfill
  \begin{subfigure}{0.49\linewidth}
    \centering
    \includegraphics[width=\linewidth]{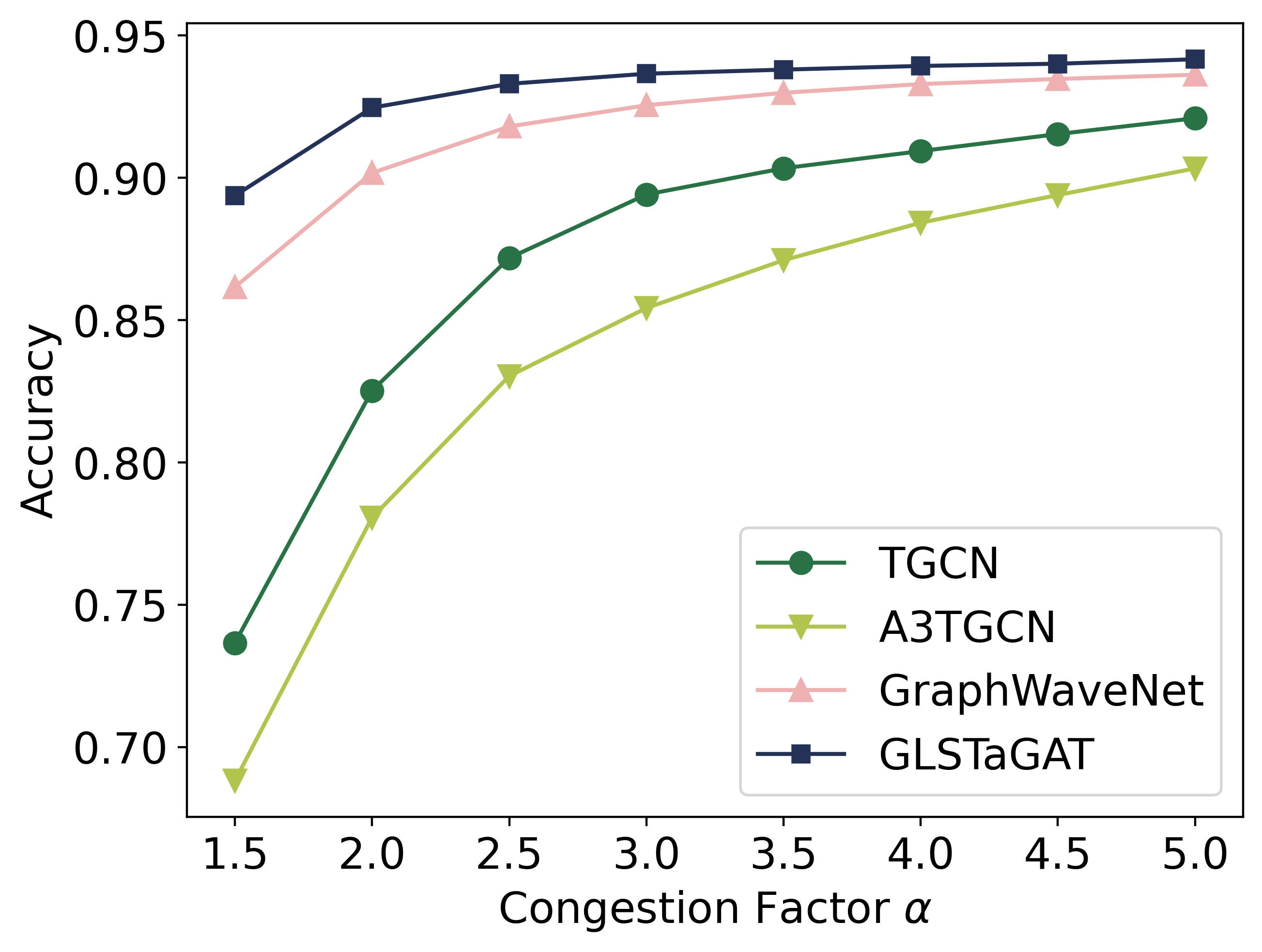}
    \caption{Comparison with Baselines (Part 2)}
    \label{accu-abilene-2}
  \end{subfigure}
  \vfill
  \begin{subfigure}{0.49\linewidth}
    \centering
    \includegraphics[width=\linewidth]{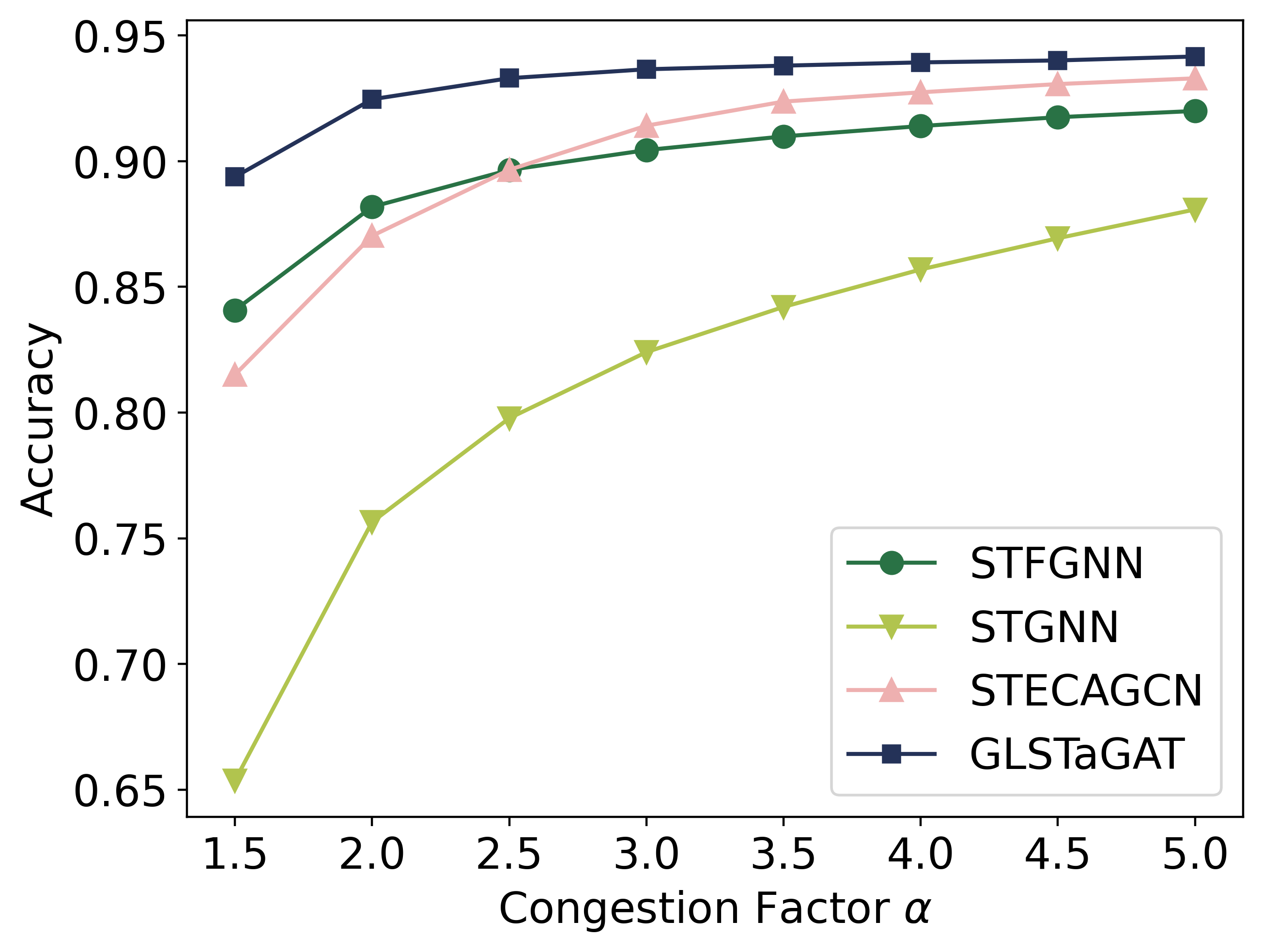}
    \caption{Comparison with Baselines (Part 3)}
    \label{accu-abilene-3}
  \end{subfigure}
  \hfill
  \begin{subfigure}{0.49\linewidth}
    \centering
    \includegraphics[width=\linewidth]{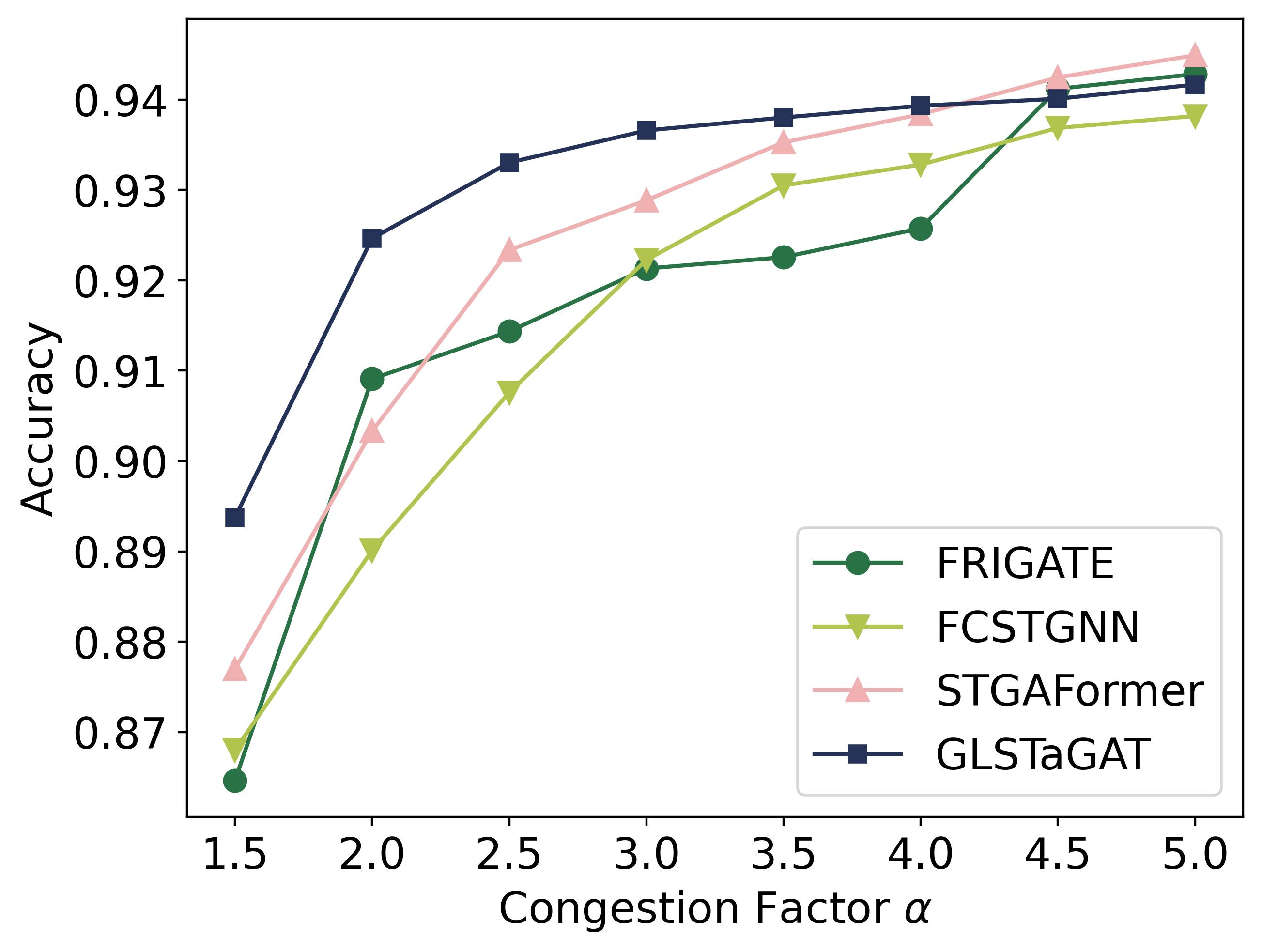}
    \caption{Comparison with Baselines (Part 4)}
    \label{accu-abilene-4}
  \end{subfigure}
  \caption{Accuracy of Detecting a Congestion on Abilene}
  \label{fig:accu-abilene}
\end{figure}

\begin{figure}[htbp]
  \centering
  \begin{subfigure}{0.49\linewidth}
    \centering
    \includegraphics[width=\linewidth]{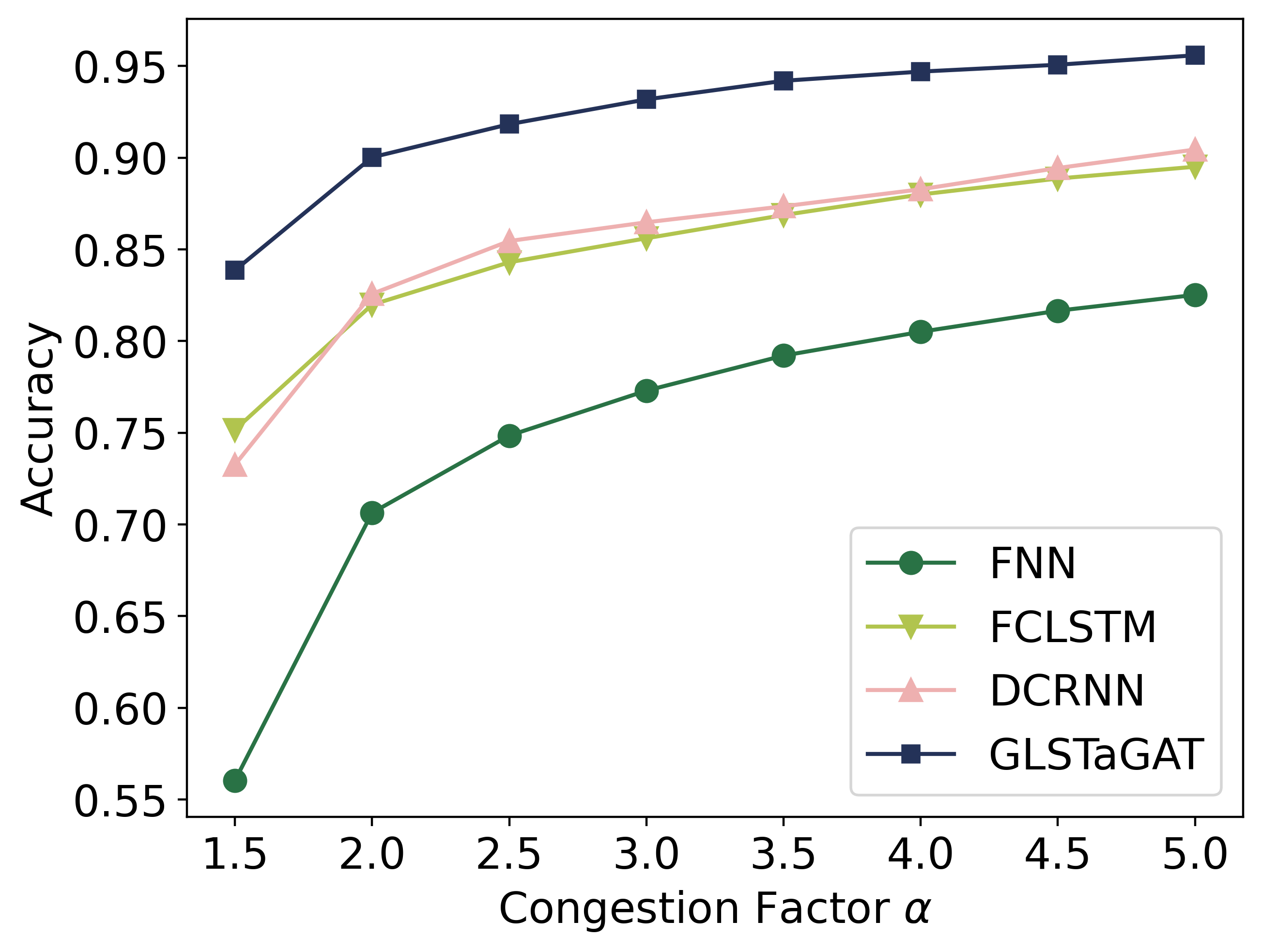}
    \caption{Comparison with Baselines (Part 1)}
    \label{accu-geant-1}
  \end{subfigure}
  \hfill
  \begin{subfigure}{0.49\linewidth}
    \centering
    \includegraphics[width=\linewidth]{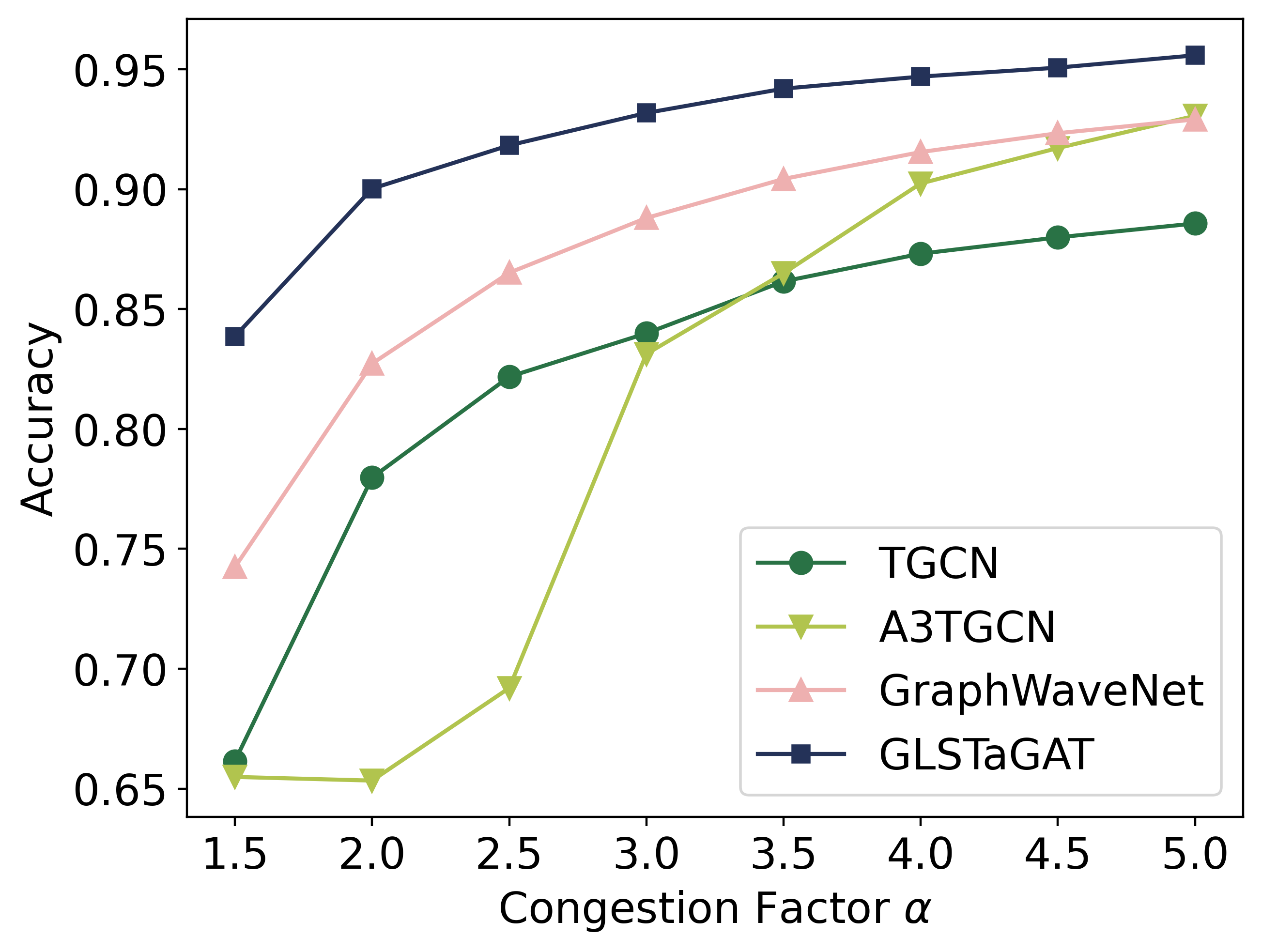}
    \caption{Comparison with Baselines (Part 2)}
    \label{accu-geant-2}
  \end{subfigure}
  \vfill
  \begin{subfigure}{0.49\linewidth}
    \centering
    \includegraphics[width=\linewidth]{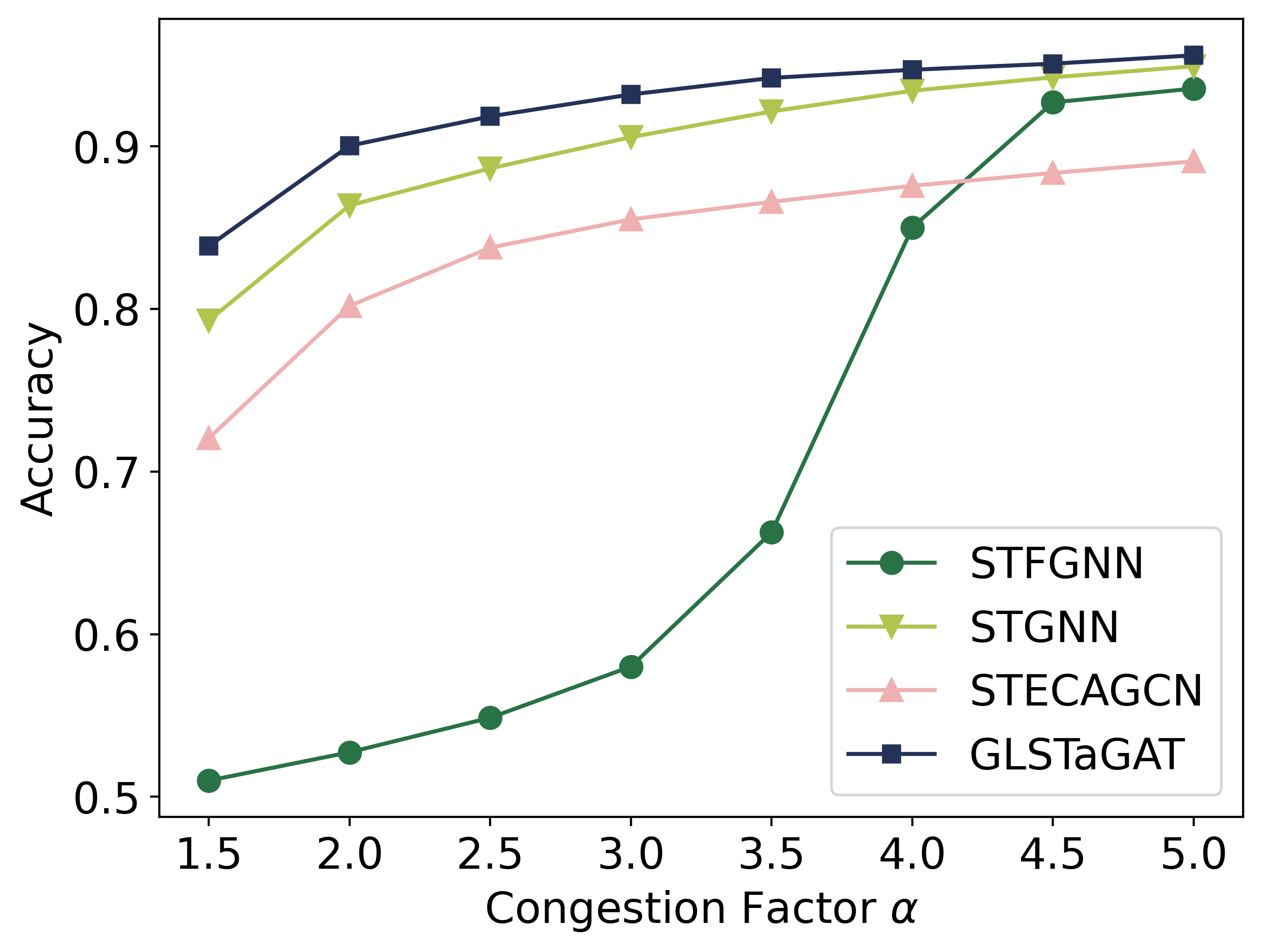}
    \caption{Comparison with Baselines (Part 3)}
    \label{accu-geant-3}
  \end{subfigure}
  \hfill
  \begin{subfigure}{0.49\linewidth}
    \centering
    \includegraphics[width=\linewidth]{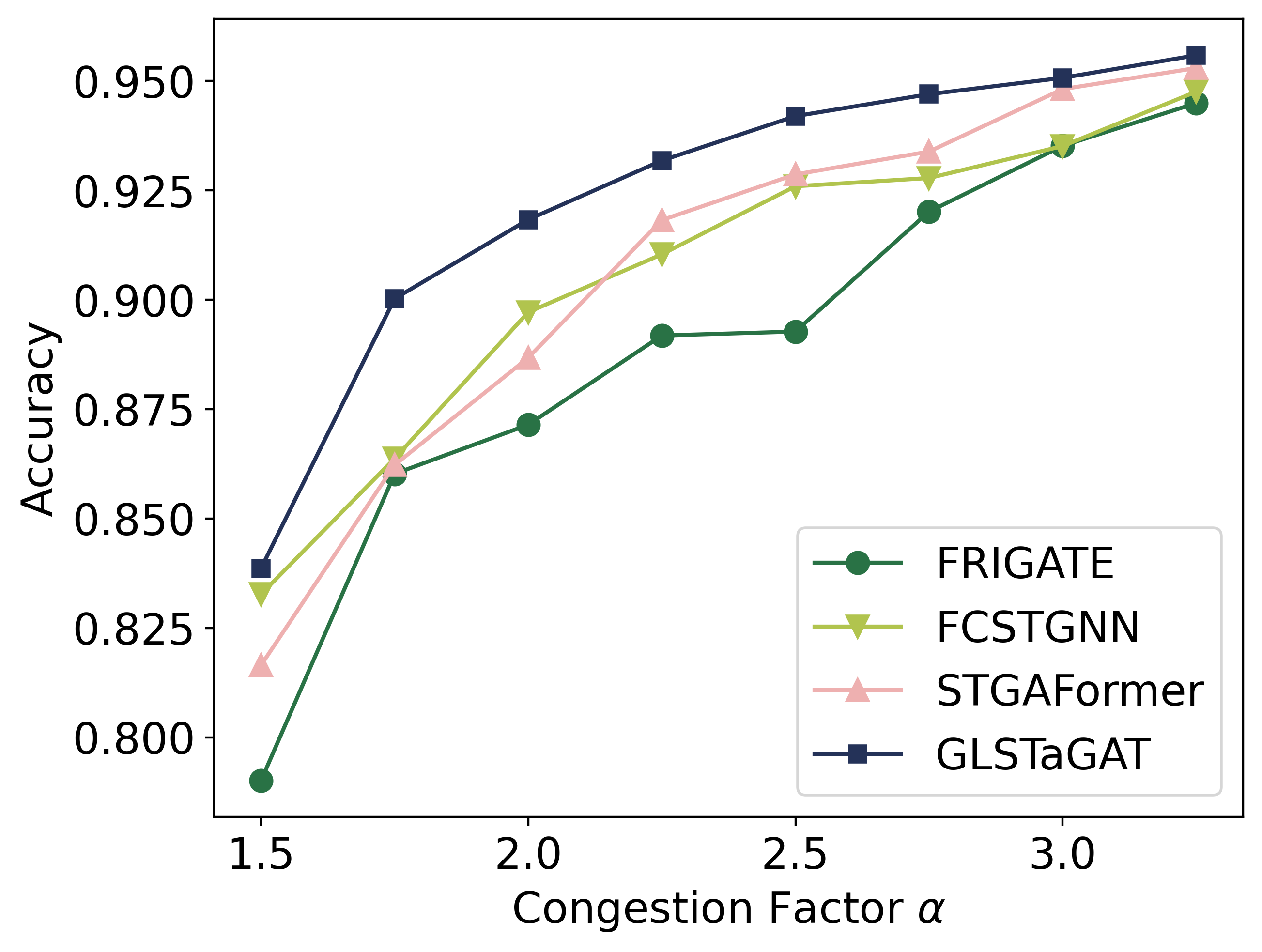}
    \caption{Comparison with Baselines (Part 4)}
    \label{accu-geant-4}
  \end{subfigure}
  \caption{Accuracy of Detecting a Congestion on GEANT}
  \label{fig:accu-geant}
\end{figure}

As indicated in \cite{andreoletti2019network}, a straightforward application of a reliable traffic predictor is congestion detection. Specifically, for each node, we assume that congestion occurs if the current traffic exceeds the congestion factor, $\alpha$, times its average traffic. In this way, we treat each node individually, successfully preserving their unique traffic patterns. The congestion factor ranges from 1.5 to 5, with a step size of 0.5. We compare the proposed model with baseline methods, and the results are shown in Figures \ref{fig:accu-abilene} and \ref{fig:accu-geant}.

\APN~outperforms the baseline models in all settings. Specifically, on the Abilene dataset, when the congestion factor is 1.5, \APN~achieves an improvement of 14.56\% compared to the average performance of the other models. As the congestion factor increases, all methods show improvement, since a larger congestion factor results in fewer congestion events. In these cases, the remaining congestion events are typically peaks, which are more challenging to predict. Nevertheless, regardless of the setting, \APN~ consistently achieves the best performance. Similar observations can be made for the GEANT dataset. These findings further indicate the effectiveness of the proposed model.
\section{Long-term Prediction}
\begin{table}[htbp]
  \centering
  \caption{Error Accumulation Analysis}
  \begin{tabular}{lccccc}
    \toprule
    SMAPE (\%) & 1T    & 2T    & 3T    & 5T    & 10T   \\
    \midrule
    Abilene    & 28.18 & 28.65 & 29.23 & 29.76 & 32.34 \\
    GEANT      & 25.25 & 27.99 & 28.55 & 28.62 & 31.59 \\
    \bottomrule
  \end{tabular}%
  \label{tab:error accumulation analysis}%
\end{table}%
To evaluate our model's robustness in long-term prediction, we conducted an error accumulation analysis. Specifically, the model was trained on data spanning time $t$ to $t+T$ and tested on future intervals of $1T$, $2T$, $3T$, $5T$, and $10T$, as presented in Table \ref{tab:error accumulation analysis}. All other experimental settings remain consistent with Section \ref{sec:ExperimentalEvaluation}. Notably, since different datasets vary in scale, SMAPE was chosen as the metric to measure relative percentage error. As shown in Table \ref{tab:error accumulation analysis}, performance declines as the prediction horizon increases due to error accumulation. Nevertheless, even at a prediction horizon of $10T$, our model maintains errors within a tolerable range, demonstrating its robustness and effectiveness for long-term prediction.
\end{document}